\documentclass[10pt,conference]{IEEEtran}

\usepackage{booktabs} % To thicken table lines
\usepackage{cite}
\usepackage[T1]{fontenc}
\usepackage{hyperref}
\usepackage{subcaption}
\usepackage{listings}
\usepackage{tabularx}
\usepackage{tcolorbox}
\tcbset{boxrule=.2mm}
\usepackage{multirow}
\usepackage{spreadtab}
\usepackage{fp}
\usepackage{todonotes}
\usepackage{comment}

\usepackage{xspace}

\newcommand{\name}{{\sc Medusa}\xspace}

\newcommand{\code}[1]{\texttt{{\small #1}}}
\newcommand{\AppName}[1]{\texttt{#1}}

\newcommand{\ie}{\textit{i.e.,}\xspace}
\newcommand{\eg}{\textit{e.g.,}\xspace}
\newcommand{\etal}{\textit{et al.}\xspace}

\newif\ifDEBUG
\DEBUGfalse

\newcommand{\jindal}[1]{\textcolor{red}{Jindal: #1}}
\newcommand{\JD}[1]{\textcolor{blue}{JD: #1}}

\usepackage{soul}
\newcommand{\TODO}[1]{\hl{#1}} % Inline highlighting
\newcommand{\qiang}[1]{\todo[author=Qiang,inline]{#1}}
\setuptodonotes{inline,noinlinepar,inlinewidth=5ex}

\ifDEBUG
\else
\renewcommand{\jindal}[1]{}
\renewcommand{\JD}[1]{}
\renewcommand{\qiang}[1]{}
\fi

\usepackage{cleveref}
\crefformat{section}{\S#2#1#3}
\crefname{figure}{Figure}{Figures}
\crefname{table}{Table}{Tables}
\crefname{listing}{Listing}{Listings}
\crefname{theorem}{Theorem}{Theorems}
\crefname{thm}{Theorem}{Theorems}
\crefname{lemma}{Lemma}{Lemmata}
\crefname{equation}{Eqt.}{Eqts.}
\crefformat{Grammar}{Grammar #1}

\usepackage{balance}

\usepackage{enumitem}

\usepackage[font=small,skip=5pt]{caption}

\usepackage{nicefrac}

\begin{document}

%\title{App Parameter Energy Profiling: Optimizing App Energy Drain by Finding Tunable App Parameters}
%\title{App Parameter Energy Profiling: Reducing Mobile Device Energy Drain by Tuning App Parameters}
%\title{Can Mobile App Re-Parameterization Decrease Energy Usage? An Empirical Study}
%\title{Searching for Free Energy: ...}
%\title{Mobile App Parameter Energy Profiling: An Empirical Study on how Parameters Impact App Energy Usage}
%\title{Mobile App Parameters Do Not Affect Energy Usage}
%\title{You Cannot Improve Mobile App Energy Performance By Tuning Parameters Without Affecting the User Experience}
%\title{If You Tune A Mobile App's Energy Performance Using Parameters, It Will Affect the User Experience}
%\title{Mobile App Parameter Energy Profiling: An Empirical Study on the Impact of Parameter Choice on Energy Usage}
%\title{An Empirical Study on the Impact of Parameters on Mobile App Energy Usage}
\title{An Empirical Study on the Impact of Deep Parameters on Mobile App Energy Usage}

\author{\IEEEauthorblockN{Qiang Xu, James C. Davis, and Y. Charlie Hu}
\IEEEauthorblockA{School of Electrical and Computer Engineering\\
Purdue University\\
West Lafayette, USA\\
Email: \{xu1201,davisjam,ychu\}@purdue.edu}
\and
\IEEEauthorblockN{Abhilash Jindal}
\IEEEauthorblockA{Department of Computer Science and Engineering\\
IIT Delhi\\
New Delhi, India\\
Email: ajindal@cse.iitd.ac.in}}

\maketitle

\newcommand{\SurveyNParticipants}{25\xspace}

\newcommand{\ExperimentNApps}{16\xspace}
\newcommand{\ExperimentNParams}{1644\xspace}
\newcommand{\ExperimentNValidParams}{2\xspace}
\newcommand{\ExperimentAvgTime}{22\xspace}

\begin{abstract}

Improving software performance through configuration parameter tuning is a common activity during software maintenance. 
%Most prior work focuses on conventional software like databases and performance metrics like latency.
Beyond traditional performance metrics like latency, mobile app developers are interested in reducing app energy usage.
%Energy usage is a key performance metric for mobile apps.
%We focus instead on mobile apps, for which energy usage is a key performance metric.
Some mobile apps have centralized locations for parameter tuning, similar to databases and operating systems, but it is common for mobile apps to have hundreds of parameters scattered around the source code.
The correlation between these ``deep'' parameters and app energy usage is unclear.
Researchers have studied the energy effects of deep parameters in specific modules, but we lack a systematic understanding of the energy impact of mobile deep parameters.

In this paper we empirically investigate this topic, combining a developer survey with systematic energy measurements.
Our motivational survey of \SurveyNParticipants Android developers suggests that developers do not understand, and largely ignore, the energy impact of deep parameters.
To assess the potential implications of this practice, we propose a deep
parameter energy profiling framework that can analyze the energy impact of deep
parameters in an app. Our framework identifies deep parameters, mutates
them based on our parameter value selection scheme, and performs reliable
energy impact analysis.
Applying the framework to \ExperimentNApps popular
Android apps, % from different categories,
we discovered that deep parameter-induced
energy inefficiency is rare.
We found only \ExperimentNValidParams out of \ExperimentNParams deep parameters for which a different value would significantly improve its app's energy efficiency. %set to a measurably problemt substantial negative set value.
A detailed analysis found that most deep parameters have either no energy impact, limited energy impact, or an energy impact only under extreme values.
%that developers can typically avoid based on their domain knowledge.
Our study suggests that
it is generally safe for developers to ignore the energy impact when choosing deep parameter values in mobile apps.

\end{abstract}

\section{Introduction} \label{sec:intro}

%%%
% Bird's-eye view
%%%

Improving energy efficiency is one of a mobile app developer's software maintenance activities. 
%Energy efficiency is an important quality attribute for mobile apps.
App users desire efficient energy usage~\cite{6682059}, and the resulting improvement in accessibility can benefit individuals and societies~\cite{MobilePhonesMatterDevelopingWorld,MobilePhoneBenefits}.
%However, it is a longstanding research challenge with major potential impact.
Mobile platform vendors, \eg Google and Apple, also advise app developers to optimize app energy usage~\cite{androidbattery,iosenergy}.

%%%
% High level of what we do
%%%
%While re-design is possible, re-configuration may be cheaper --- provided the app developers know which configuration parameters to tune.
One potential strategy to reduce a mobile app's energy usage is to tune its configuration parameters.
All software includes configuration parameters to help it be adapted to different contexts.
Mobile apps are no exception: in addition to the parameters explicitly exposed in resource files and other configuration files, these apps have many \emph{deep parameters}, \ie parameters that are scattered around the source code to control runtime behaviors including buffer sizes, task frequencies, and UI layout positions.
Researchers have shown that tuning parameters can improve the performance of conventional software~\cite{10.1145/3127479.3128605,10.1145/3035918.3064029,10.14778/3352063.3352129}.
However, mobile deep parameters are often overlooked by developers, and little is known about the energy impacts of deep parameters in mobile apps.
Prior works have only studied the energy impacts of deep parameters in specific modules~\cite{10.1145/3236024.3236076} or libraries~\cite{10.1145/3067695.3082519,10.1145/3286978.3287014}, not systematically. % understanding of the energy impact of mobile app parameters.
%lead to app may render the map unnecessarily frequently and induce a high energy drain when the location is updated too frequently.
%Incorrectly configured parameters can cause unnecessarily high app energy usage.
% In this paper, we focus on the parameters' energy effects, as battery life is a big resource constraint for smartphones.

%%%
% Overview - Survey
%%%

We investigated the energy impact of mobile deep parameters using mixed methods~\cite{johnson2004mixed}, combining a developer survey with systematic energy measurements. In our survey of \SurveyNParticipants Android app developers, we found that developers are uncertain about the energy impact of deep parameters and do not usually consider energy when choosing parameter values.
%JD: I fear the next sentence will disappoint the reader with our negative result.
%This may result in energy inefficiency if parameters are poorly configured in terms of energy drain.

%%%
% Overview - Systematic measurement 
%%%
To measure the implications of developers' practices on deep parameters, we propose a parameter-centric energy profiling framework.
The framework extracts deep parameters from the app, mutates them based on our parameter value selection scheme, and measures the changes in energy drain.
Our framework overcomes several challenges:
   identifying deep parameters,
   choosing appropriate mutation values,
   and reliably measuring the energy impact.
%Furthermore, automated tests need to be carefully designed when repeated a large number of times.

We systematically measured deep parameter energy effects in \ExperimentNApps popular open-source Android apps. Among the \ExperimentNParams parameters tested, only \ExperimentNValidParams are set to energy-inefficient values. Further analysis shows that the rest of the parameters either have no energy effect, have limited energy effects, or only have energy effects under extreme values that developers can typically avoid based on their domain knowledge.
We conclude that it is generally safe for developers to ignore energy effects when choosing deep parameter values --- developers must look elsewhere for energy-reducing refactorings.

Our study makes the following contributions:
\begin{itemize}
\item We describe the practices of mobile app developers on deep parameters and energy optimization (N=\SurveyNParticipants) (\cref{sec:DevSurvey}).
\item We propose a framework for parameter-centric profiling in Android, automatically identifying deep parameters and measuring their energy impacts (\cref{sec:framework}).
\item We perform the first systematic study of the energy impact of deep parameters in Android apps (N=\ExperimentNApps) (\cref{sec:results}). We describe the roles of deep parameters in these apps, and identify three energy categories of deep parameters.
\item We open-source our framework
	and full survey and experiment results%
	\footnote{\urlstyle{tt}\url{https://doi.org/10.5281/zenodo.5823364}}
	to enable reproducibility and further exploration from the research community.
\end{itemize}

\section{Background and Definitions}

\subsection{Configuration Tuning in General}

Many categories of software can be configured for different usage scenarios and deployment environments. Such software includes databases, stream
processing frameworks, web servers, codecs, and others. Their
configuration parameters are typically exposed through configuration files,
command-line interfaces, or certain data
structures~\cite{10.1145/3127479.3128605}. For example, to configure the video
codec \texttt{x265}, one can pass command-line arguments to the
executable~\cite{x265cli}, or specify the \texttt{x265\_param} data
structure through its API~\cite{x265api}.

In addition to their implications on software functionality, configuration
parameters may also impact performance metrics.
For example, by tuning its $\sim$200 configuration parameters, MySQL database throughput can be improved by 6x and its latency reduced by 3x on common
benchmarks~\cite{10.14778/3352063.3352129}.
%The parameters controlling various buffer and cache sizes are the most important ones~\cite{10.1145/3035918.3064029}.
As configuration tuning is an NP-hard problem~\cite{10.1145/1005686.1005739}, configuration tuners aim to efficiently search the configuration space and recommend configuration values for a given workload (\eg~\cite{10.1145/3127479.3128605,10.1145/3035918.3064029}).
In these contexts, auto-tuners are able to focus on the performance optimization task because the software parameters are clearly defined (\eg in configuration files).
%Such configuration tuners can outperform manual expert tuning. %achieve better performance than manual tuning performed by experts.

\subsection{Parameters in Mobile Apps}
\label{subsec:mobile-param}

% Pandora: view updates
% Spotify: method count or GfxDoctor
% OsmAnd: GfxDoctor
% Slide: wasted computation and network

Mobile apps also contain configurable parameters that control various aspects
of the apps.
%Common parameter types include upper/lower bounds, UI layout sizes and positions, buffer/cache sizes, thread counts, timeouts, and task frequencies.
As in most user-facing software, latency is a major performance metric in mobile apps.
However, \emph{energy} is also a key metric in mobile apps.
Similar to configuration tuning for other performance metrics, other researchers have provided preliminary evidence that some parameters impact app energy consumption.
Canino \etal~\cite{10.1145/3236024.3236076} showed that GPS configuration parameters can be tuned to meet specified energy consumption SLAs.
Similarly, Bokhari \etal~\cite{10.1145/3067695.3082519,10.1145/3286978.3287014} optimized the energy consumption of the Rebound physics library by tuning its numeric parameters.
However, it is unclear whether such energy-affecting parameters are common in general Android apps.

This problem is challenging because, in contrast to the
software discussed above, 
% we report that 
parameters in Android apps
are more often scattered all around the source code rather than stored
at central places.  We speculate that parameter centralization occurs
when the software is designed to be maintained by skilled operators such as database
administrators.  Most mobile apps are designed for unskilled users,
and so there is little customer demand to centralize and expose
parameters.  This challenge must be overcome in order to understand
the impact of parameters on mobile app energy usage.

\subsection{Definitions} \label{sec:bg-definitions}

\begin{figure}
\vspace{-2mm}
       {\small
\lstset{numbers=left,xleftmargin=3.5ex,moredelim=[is][\color{red}]{|}{|}}
         \begin{lstlisting}
Bitmap.createBitmap(|320|, |240|, |ARGB_8888|);
byte[] serverVersion = new byte[|512|];
sock.setSocketTimeout(|0|);
layoutParams.width = |12|;
         \end{lstlisting}
         }
\vspace{-2mm}
\caption{
  Real Android code snippets; deep parameters in \color{red}{red}.
  }
\label{fig:param}
\vspace{-4mm}
\end{figure}

Following Bokhari \etal~\cite{10.1145/3067695.3082519,10.1145/3286978.3287014}, we define a \textbf{deep configuration parameter} as \emph{a constant in app source code that can be changed by app developers, but does not affect app functionality}.
In other words, all app components should function properly when tuning a deep parameter's value, with ``minimal'' impact on user experience.
%This includes all numeric constants, Boolean constants, and enumerator references ().
This property can be determined by examining the source code or the runtime app behavior.\footnote{A high-quality test suite would be a suitable oracle, but we found these suites inadequate in our experiments.}
%For example, values representing failure types should not be changed arbitrarily, as otherwise the app may report the wrong kind of failure.
%Constants that only function properly with a limited range of values are also considered parameters.
Some examples are given in~\cref{fig:param}, \eg buffer sizes, timeouts, and UI element sizes, and the parameters can be numeric, Boolean, or enumeration values.

\section{Research Questions}

We seek to understand the energy efficiency of Android deep parameters. 
%We study the current practices of Android developers, as well as experiment with parameters in existing apps.
App energy efficiency might be affected by developers' awareness of the energy impact of deep parameters and their strategy in deciding parameter values.
We also want to measure and understand the energy effects of tuning deep parameters.
Operationalized, our research questions are:

\vspace{0.1cm}

\begin{itemize}[nosep,leftmargin=0.4in]
\item[\textbf{RQ1:}] What are the energy impacts of parameters in developers' eyes?
\item[\textbf{RQ2:}] How do developers choose parameter values?
\item[\textbf{RQ3:}] Is deep parameter-induced energy inefficiency common among apps?
\item[\textbf{RQ4:}] When and why do (and do not) deep parameters impact app energy consumption?
\end{itemize}

\vspace{0.1cm}
\noindent
We investigated RQ1-2 with a developer survey.
We studied RQ3-4 through a systematic energy measuring experiment.

\section{Developer Perspectives} \label{sec:DevSurvey}

We surveyed Android developers to better understand their perceptions and
practices regarding Android parameters.

\subsection{Methodology}

We designed an IRB-approved survey to obtain mobile app developers' perspectives on RQ1 and RQ2.
The RQs were operationalized into 6 demographic questions and 13 study-specific questions.
These questions included closed- and open-ended questions across three topics:
  (1) the nature of the parameters in their apps;
  (2) their perceptions of parameters' energy impacts;
  and
  (3) their process when choosing parameter values.
Rather than using interviews to elicit relevant topics for the survey, we based the questions on
  (a) our own expertise from studying and developing energy-efficient mobile apps;
  and
  (b) preliminary findings and observations from our energy experiments (detailed in~\cref{sec:framework,sec:ExperimentalDesign,sec:results}).
%To ensure our survey questions are valid and not biased,
To improve instrument validity and reduce bias,
  we followed best practices during survey design~\cite{pew,questionpro}, \eg avoiding leading questions, and refined the survey through two rounds of pilot studies with graduate students.
%\JD{It is conventional to prefix a survey with interviews or focus groups to refine the topics. We did not do so because of the team's expertise in experiments.}

To ensure participant's understanding of the term ``parameter'', we provided the following definition before survey questions:
  ``\textit{Android apps contain parameters that control various aspects of the apps. Common types of parameters include upper/lower bounds, UI layout sizes/positions, buffer/cache sizes, thread counts, timeouts, task frequencies, etc. They could be hard-coded numbers, constants, or dynamically varying. In this survey, we are interested in the parameters that are accessible to developers, i.e., not in-app settings.}''\footnote{This definition includes both deep (cf. \cref{sec:bg-definitions}) and traditional parameters. Our results thus give a broader perspective on the energy tuning practices of mobile app developers. As deep parameters are a subset of this definition, our survey results also characterize engineering practices for deep parameters.}
Respondents were asked to answer in terms of the app they spent the most time developing. %, in case they have worked on multiple apps.

We distributed the survey to Android developers through multiple channels.
We posted the survey on
  popular forums
    frequented by Android developers (subreddits \texttt{r/androiddev} and \texttt{r/mAndroidDev})
    and
    developers in general (Hacker News and DEV),
  and social media groups of Android developers (LinkedIn, Facebook, and Slack).
We also contacted Android developers in our professional networks.
Survey participants were not compensated.

We received 25 non-blank responses:
  15 from forums and social media groups,
  and
  10 from professional connections.

\vspace{-0.2cm}
\subsection{Results}

\paragraph{Demographics}
The median respondent has 6-10 years of software development experience and 3-5 years of Android development experience.
Respondents work on apps from \nicefrac{13}{37} categories defined by Google Play~\cite{googleplay}.
For \nicefrac{21}{25} responses, the answers describe commercial app development.

\begin{figure}
	\subcaptionbox{
	    For what proportion of the parameters in your app are you confident about the energy impact of changing them?\label{fig:survey-confidence}
	}
	{\includegraphics[width=\columnwidth,trim=0 15 0 15]{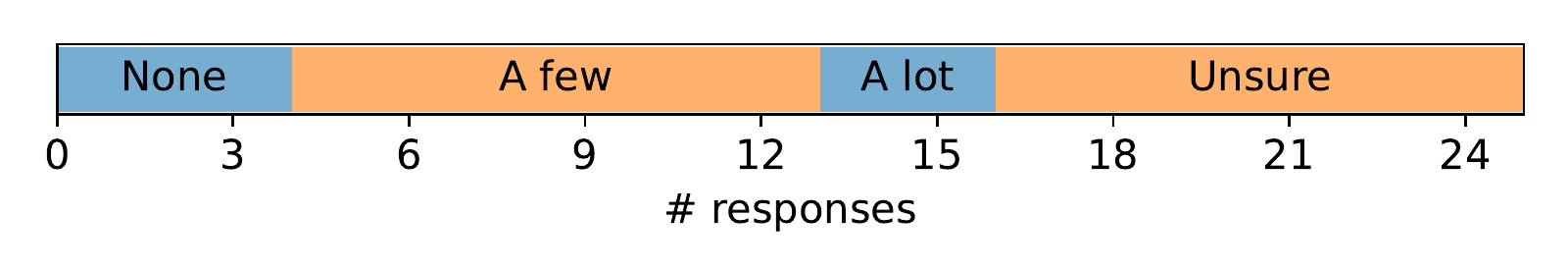}}
	\subcaptionbox{
	  How often do you consider energy consumption while choosing parameter values? \label{fig:survey-pick-value}
	}
	{\includegraphics[width=\columnwidth,trim=0 15 0 0]{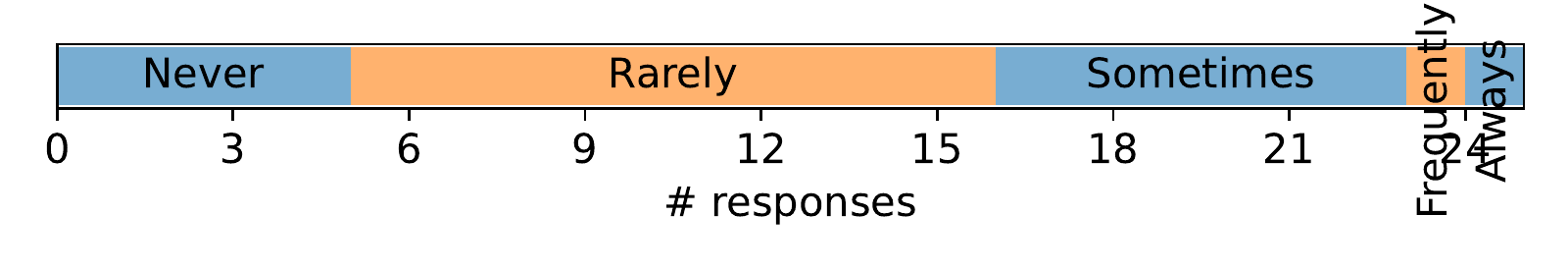}}
	\caption{
    	Distribution of responses to developer perception of parameters' energy impact and how they pick parameter values
	}
	\vspace{-4mm}
\end{figure}

\paragraph{RQ1: Developer Perception of Energy Impact}
Developers are concerned about app energy usage: 10 of the 25 respondents monitor energy consumption.
However, most of these respondents use coarse-grained measurements like CPU usage or battery statistics. % in Settings, CPU usage, or Android Studio Battery Profiler.
These tools can detect severe or specific types of energy bugs (\eg wake lock~\cite{10.1145/2950290.2950297}), but are difficult to use for energy tuning.
Perhaps in consequence, only 3 respondents are confident about the energy impacts of ``a lot'' of their apps' parameters (\cref{fig:survey-confidence}).
%We conjecture the lack of good tools for measuring energy usage to be one contributing factor.

\begin{tcolorbox}
    \textbf{Finding 1}:
    Around half of mobile app developers measure app energy usage.
	Few developers (12\%) are confident about the energy impact of
	parameters. 
\end{tcolorbox}

\paragraph{RQ2: Picking Parameter Values}
Our respondents said that when they choose parameter values, their top concerns are app functionality and user experience.
Only 2 of the 25 developers frequently or always consider energy consumption when parameterizing (\cref{fig:survey-pick-value}).
The reason might again lie in the fact that developers don't have handy tools for energy tuning.
As developers have only limited confidence in parameters' energy impacts, further experiments are still needed to validate developers' choices.

\begin{tcolorbox}
    \textbf{Finding 2}:
	Only 8\% of developers frequently consider energy consumption when
	choosing parameter values.
\end{tcolorbox}

\paragraph{Parameter Locations}

Other data from our survey informed our parameter measurement approach.
For parameters in source files, our respondents estimated that the majority are scattered across the codebase; only a third of respondents described their apps as having substantial parameter centralization in files like \texttt{Config.java} or \texttt{Constants.java}.
This finding is consistent with our observations of open-source Android apps, discussed in~\cref{subsec:mobile-param}.
%The survey corroborates the observation for commercial apps.

\iffalse
\begin{figure}
	\includegraphics[width=\columnwidth]{fig/survey-param-location}
	\caption{Where are the parameters stored in your app?}
	\label{fig:param-location}
\end{figure}
\fi

\section{Deep Parameter Testing Framework} \label{sec:framework}

\begin{figure*}
\includegraphics[width=\textwidth]{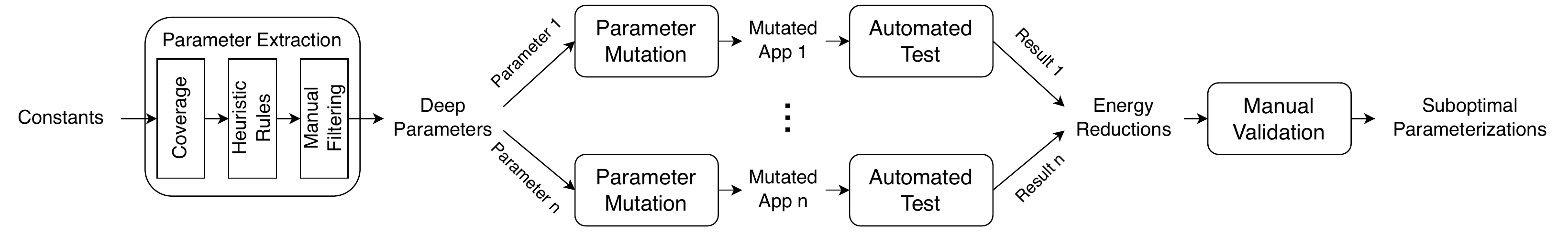}
\caption{Workflow of the mutate-and-test process.}
\label{fig:workflow}
\vspace{-4mm}
\end{figure*}

The developers in our survey indicated that they rarely consider energy consumption when picking deep parameter values.
This practice does not necessarily mean that they make poor choices.
Developers might intuitively make energy-efficient choices; energy-efficient choices may correlate with choices that improve usability; or deep parameters may not have a substantial effect on energy usage.
%Developers should change their practices if suboptimal parameterizations are common.
However, since developers told us that they do not consider the energy effects of parameters, our governing hypothesis is that \emph{ignoring energy effect results in suboptimal deep parameterizations}.
%\JD{For reasons discussed in~\cref{sec:framework-alternativeDesign}, our framework opts for a conservative approach}
To test the hypothesis, we propose a framework that mutates every deep parameter and checks if the change reduces energy usage.

An overview of the mutate-and-test process is
shown in Figure~\ref{fig:workflow}.
As deep parameters are scattered in the source code, we first extract deep parameters from the set of all constants. For each of the parameters extracted, we
try several new values based on our parameter mutation scheme, and measure the
energy consumption of the app variants. We manually validate all parameterizations that
reduce energy use, and finally report any discovered energy-reducing
parameters.

We present the details of each step below.
Many design details are informed by preliminary experiments and observations.
The primary design constraint is \emph{time}: testing one parameter takes a long time (on average \ExperimentAvgTime minutes), as we need to perform repeated tests for multiple parameter values on the phone; and these tests are not easily scaled, since we need a physical device to get accurate energy measurements.

We describe our experiment applying the framework to \ExperimentNApps popular Android apps in \cref{sec:ExperimentalDesign} and the results in \cref{sec:results}.

\subsection{Deep Parameter Extraction}

Deep parameters are scattered around source files.
Thus, to test the parameters, one
design option is to test all constants, regardless they are deep parameters or not. However, an app has thousands of constants and testing all of them is prohibitively time-consuming.
We therefore use a mix of automatic and manual filtering to distinguish deep parameters from other constants.
%To keep the test time manageable, we extract deep parameters scattered around by filtering out the constants that are not used or not deep parameters using a combination of automatic and manual filtering, discussed below.

\paragraph{Usage Scenarios and Parameter Coverage}
Our framework tests an app using UI automation scripts over several of its usage scenarios.
After we identify these usage scenarios, we use code coverage to identify the candidate deep parameters.
A usage scenario may target a particular feature set (\eg features to be used in a low-power mode), and only some deep parameters will affect the energy usage in this scenario.
Deep parameters that are not used cannot have an energy effect.
Based on this insight, we record the line-level code statement coverage and filter out parameters that are not covered.
This filtering is performed prior to the energy measurements, thus energy measurements are conducted on a non-instrumented app and do not have instrumentation effects.
%Filtering parameters via coverage analysis is expensive, but is only performed once for each use scenario.
%For the actual energy measurement tests, we run the non-instrumented app, without paying the coverage performance penalty and concomitant energy consumption.

\paragraph{Heuristic Rules}
We noticed that true deep parameters are more likely to occur in certain code constructs, while constants in some other code constructs are unlikely to be deep parameters.
For example, variable initializers and method call arguments tend to be deep parameters; bitwise operator arguments typically not.

Using ``negative patterns'' matching non-parameter constants minimize the chance that we incorrectly filter out deep parameters. Thus, we assemble a set of patterns that are shared by portions of the non-parameter constants, and filter
out constants that match any pattern. The patterns are extracted by
manually classifying constants in a file to be deep parameters or not, and identifying any patterns that the non-parameter constants may share.

After manually inspecting the files with the most constants, we derived heuristic rules as shown in~\cref{tab:rule}.
Each rule may be applicable to one or several parameter types.
As the rules are more pertinent to programming language idioms~\cite{10.1145/2635868.2635901} than an individual developer's habits, they apply across apps.
However, as the rules depend on common development practices, they may incorrectly filter parameters (false negatives).

\begin{table}
\caption{Heuristic rules for filtering non-parameter constants.}
\label{tab:rule}
\small
\centering
\begin{tabular}{cll}
\toprule
Type  & Rule & Example \\
\midrule
Num & Array index & a[0] \\
Num & Comparison with 0 or 1 & a.size() > 0 \\
Num & Plus 1 or 2 or minus 1 & a.length() - 1 \\
Num & For loop initialization & for (int i = 0; ...) \\
Num & Ignored methods & s.substring(0, 4) \\
Bool & One argument method call & item.setVisible(true) \\
Enum & Time unit & convert(5, DAYS) \\
Enum & Locale & toLowerCase(US, str) \\
All & Condition & if (a.size() > 0) \\
All & Return value & return 0 \\
All & Multiple writes to variable & See \cref{fig:multi-write} \\
\bottomrule
\end{tabular}
\vspace{-2mm}
\end{table}

\paragraph{Manual Filtering}
The preceding filters are automated. %coverage- and heuristic-based filtering are performed automatically.
However, many constants remain for consideration.
%But as our heuristic rules do not cover all the non-parameter constants, we still have to test more constants than needed.
We manually examine the remainder and filter out non-parameter constants based on their semantic.
For example, error codes are not parameters, since mutating them will make the app report the wrong kind of error, and thus affecting app functionality.
%As automatic filtering has already filtered out most of the non-parameter constants, we only need to examine a small fraction of the constants.

\subsection{Deep Parameter Mutation}

\begin{figure}
  {\small
    \begin{lstlisting}[language=Java]
class Foo {
	int counter = 0;
	void count() { counter++; }
	void reset() { counter = 0; }
}
\end{lstlisting}
\vspace{-2mm}
\caption{
  Example of multiple writes.
  \texttt{counter} is updated in multiple places, so the \code{0} constants are not considered deep parameters.
}
\label{fig:multi-write}
\vspace{-4mm}
  }
\end{figure}

We mutate and test each deep parameter in isolation.
It is intractable to test all possible values for each parameter, thus sampling is used for value selection. We choose
new values carefully to maximize the chance that we observe a parameter's
energy impact. For example, the energy consumption may not change if we choose
values too close to the original one, yet we may crash the app if we choose
values too far away. On other hand, the new values do not need to be optimal,
as we can perform further investigation as long as we can observe an energy
reduction.

Based on preliminary experiments, we developed a guideline for choosing new values for numeric parameters:
\begin{itemize}
	\item Choose values from both sides of the original value; most
		parameters are monotonic in terms of their energy effects.
	\item A factor of 8 will expose the difference (if
	      any) in energy consumption, unless the parameter only has energy impact under
	      extreme cases (cf. \S\ref{subsec:extreme}).
\end{itemize}

We further fine-tune the values chosen based on the original parameter value, as we observed that the original value indicates the valid value range to some
extent. For example, 0 is invalid for many positive integer parameters, while
floating-point parameters with original values between 0 and 1 are highly
likely to be valid only between 0 and 1.

Combining the guidelines and the fine-tuning, we build a parameter mutation
scheme for numeric parameters as shown in Table~\ref{tab:new-value}. Integer
parameters with original value 0 have versatile semantics and do not fit into
our guidelines. Thus, we choose several special values commonly used in
programming to maximize the chance that some values suit the semantics. %fit the parameter.

\begin{table}
\caption{
  Mutated values for numeric parameters. % based on the original value.
  }
\label{tab:new-value}
\centering
      {\small
        \begin{tabular}{cc}
\toprule
Original value & New values \\
\midrule
$0$ & 0xffffff, $255$, $8$ \\
$1$ & $8$, $0$ \\
$x$ ($> 1$, int) & $x * 8$, $\max(x / 8, 1)$ \\
$0.0$ & $0.5$, $1.0$ \\
$x$ ($0 < x < 1$, float) & $1 - (1 - x) / 8$, $x / 8$ \\
$x$ ($\ge 1$, float) & $x * 8$, $x / 8$ \\
\bottomrule
\end{tabular}
      }
\vspace{-2mm}
\end{table}

Boolean and enumerator parameters are simpler than numeric parameters.
We invert Boolean parameters, and for enumerations we randomly choose three additional values.

\subsection{Automated Testing}

The high-level workflow of automated testing is simple: we drive each app with
a deterministic UI automation script. For each deep parameter, we measure the energy
drain for both the unmodified app and the app with a new parameter value, and
finally compare the results to see if the parameter can reduce the app energy
drain. However, to ensure that the tests are reproducible, statistically solid,
having minimal false positives and false negatives, and faithfully reflect the
effects of
the parameters, every step needs to be carefully thought out.

\subsubsection{UI Automation Script}

There has been a large body of research on automated UI testing for Android apps~\cite{8952363,9152712,10.1145/3377811.3380347,10.1145/3377811.3380402,10.1145/3377811.3380382,10.1145/3395363.3397354}.
However, these works aim at improved code coverage while our automated testing instead focuses on reproducibility.

We design one UI automation script for each test scenario. While it is
relatively easy to write a script that
% works when you 
runs for a couple of
times, extra caution is needed to design a script that runs thousands of
times and ensures everything is reproducible at the same time.
%In addition to obvious considerations such as cleaning caches, there are more subtle considerations. 
We enumerate four lessons we learned from our experience.
%  for the research community:

% when to start measurement
% cache

% same operations: amaze
\paragraph{Ensure the interactions are deterministic}
Executing the same script each time does not guarantee that the app performs
the same actions.
For example, the problematic code in~\cref{fig:determ} swipes on a scrollable list until it  
finds the desired item.
If the list contents or order vary, the app behavior will also vary, affecting energy usage.
%The interactions become non-deterministic if any of the conditions are not met.
Avoid such loops with non-deterministic terminating conditions.

% reproducible content: Slide, several local apps
\paragraph{Ensure the test data are also deterministic}

App behaviors depend on both the interactions and the data fed into the apps.
Watching different videos or reading different posts may consume different
amounts of energy. This issue is easy to solve for local apps like galleries or
file managers, but harder for apps that rely on remote content.
For example, the most popular posts on Reddit change.
We address this by accessing static content whenever possible. %keeping the accessed content as static as possible.
In the case of the Reddit client \AppName{slide} (cf. \cref{tab:app}), instead of fetching trending posts, we fetch the most popular posts of all time.

% saving bandwidth, saving time: newpipe, osmand, kpd, ap
\paragraph{Save app data to save bandwidth and time}

Many apps need to download a large amount of data when opened for the first
time. For example, \AppName{fdroid} needs to download tens of
megabytes of metadata for all apps in the store. While this is acceptable for a
small test, downloading the data several hundred times can easily lead to
protective measures like reduced bandwidth or even blockage on the server-side.

On the other hand, preparing an app for a test may be very time-consuming.
For example, to realistically test a password manager app, the app's database should have dozens of password entries.
However, popping the database with so many entries for every
fresh install takes time. % and slows down the test.

The solution for both scenarios is to utilize the data export functionality
provided by many apps. By exporting the bootstrap data, all subsequent tests
only need to import them locally after app installation.

% pay attention to server state: conv
\paragraph{Pay attention to the server state}

For many apps requiring account login, part of the state is stored on the
server-side. Without resetting the server state, the app may behave non-deterministically. For example, the server of the instant messenger
\AppName{conv} recognizes the phone as a new device every
time the app reinstalls. In the end, the server maintains thousands of
``devices'', which drastically changes the app behavior. As servers are black
boxes, this type of problem is hard to diagnose. Our solutions were  
app-specific.

\subsubsection{Back-to-Back Testing}

\begin{figure}
\vspace{-2mm}
  {\small
    \begin{lstlisting}[language=Java]
while (!onScreen(item))
	swipe();
\end{lstlisting}
  }
  \vspace{-2mm}
  \caption{UI automation code with potentially non-deterministic test interactions. This could lead to flaky test results.}
\label{fig:determ}
\end{figure}

\begin{figure}
  \subcaptionbox{Energy consumption over a period of 3 days. The error bar represents the standard deviation of the 5 runs.
    \label{fig:ref-energy-time}}
{\includegraphics[width=0.95\columnwidth,trim=0 10 0 0]{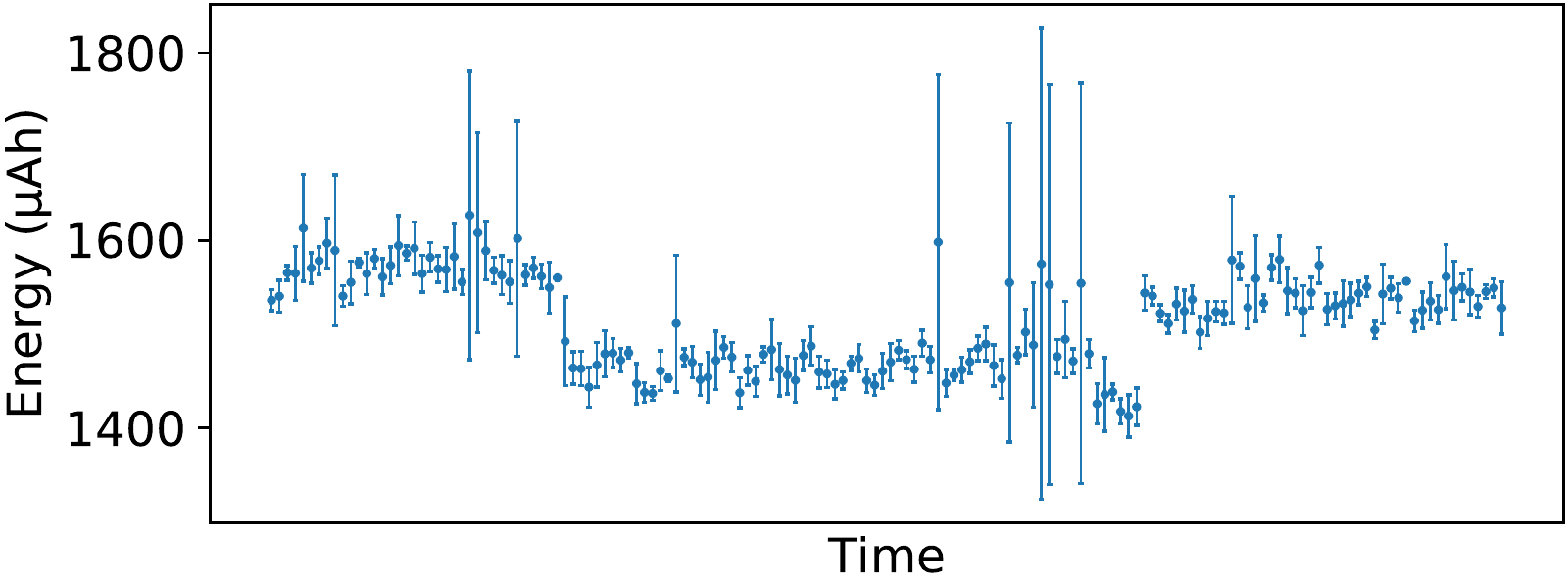}}
\subcaptionbox{The CDF of the normalized standard deviations in \cref{fig:ref-energy-time}.
  \label{fig:ref-sd-cdf}}
{\includegraphics[width=0.95\columnwidth,trim=0 10 0 0]{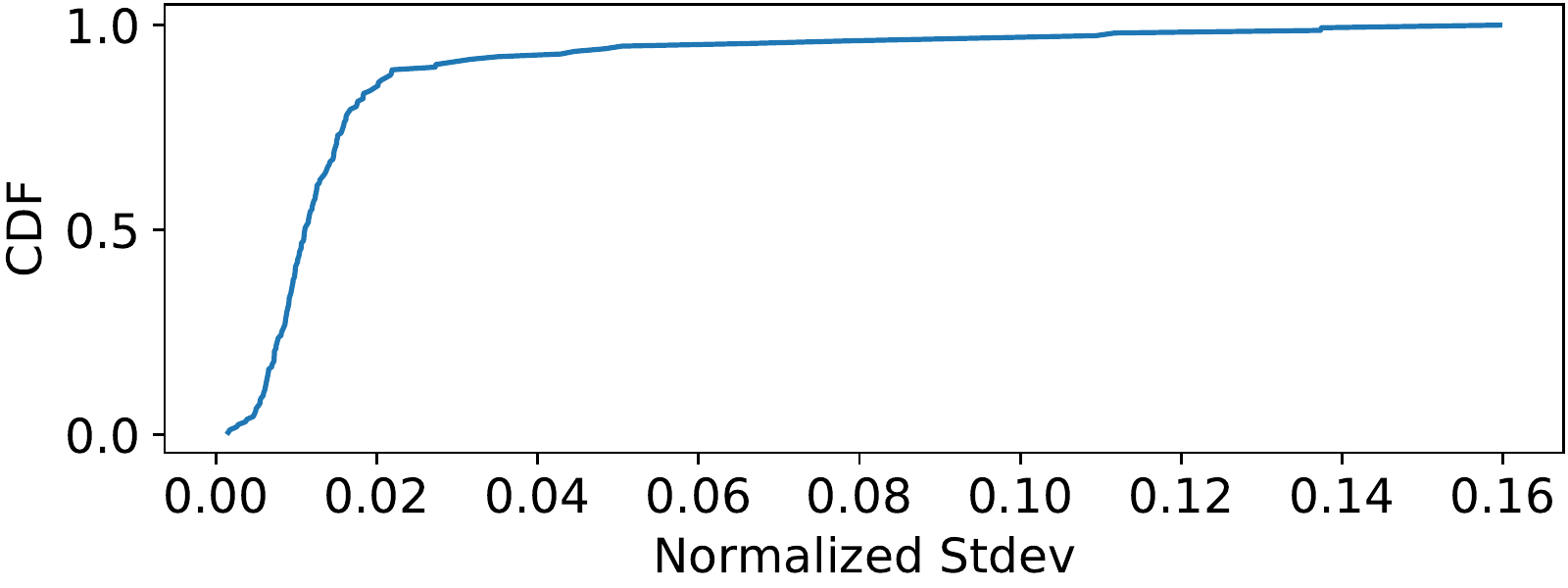}}
\caption{Energy consumption of the unmodified \AppName{ap} app. Each data point
	represents 5 runs.}
\vspace{-4mm}
\end{figure}

Due to the time-dependent network condition and server load, the app energy
consumption is changing over time for many apps with network access.
Figure~\ref{fig:ref-energy-time} shows the energy consumption of \AppName{ap} over a period of 3 days. While adjacent data points typically have
similar energy consumption, the energy consumptions of distant data
points can differ by as much as 14\%.
To reduce the influence of such time-dependent energy consumption drifts, we
rerun the unmodified app before testing each deep parameter, and only compare the
adjacent tests.

\subsubsection{Hypothesis Testing}

To determine whether a parameter value reduces energy consumption, we run both
the unmodified app and the modified app 5 times (denoted as $B_1, \ldots, B_5$
and $P_1, \ldots, P_5$ respectively), and perform Student's $t$-test with the
null hypothesis being mean($P_i$) = mean($B_i$) and the alternative hypothesis
being mean($P_i$) < mean($B_i$), and a significance level of 0.05.

However, when applying the above hypothesis testing, we noticed that many
energy drain reductions are due to small energy consumption fluctuations,
rather than the parameters. To reduce such false positives, we set a threshold
$t_d$ for each app and filter out energy differences smaller than the
threshold. Specifically, instead of performing the $t$-test for
$P_i$ and $B_i$, we now perform t-test for $P_i$ and $B'_i$, where
$B'_i = (1 - t_d) B_i$. Since $P_i$ and $B'_i$ have different variances, we
switch from Student's t-test to Welch's t-test.

\subsubsection{Stability Threshold}

In the last section, we addressed the false positives caused by energy
fluctuations. However, energy fluctuations can also introduce false
negatives.
\cref{fig:ref-sd-cdf} plots the CDF of the normalized standard deviations ($\sigma$/$\mu$) using the same data as~\cref{fig:ref-energy-time}.
Although the standard deviation is low ($<3$\%) in most cases, intermittently it can reach  
16\% of the mean. The chance of passing the $t$-test is minimal with such
a high standard deviation.
In such cases, we choose another small threshold $t_s$ 
for each app, and discard results until the normalized standard deviation is back to normal (less than $t_s$).
Both $t_d$ and $t_s$ are determined through experiments (cf. \cref{sec:ExperimentalDesign}).

\subsection{Manual Validation}

In the last step,
we manually validate the energy-reducing cases to make sure that they are
really caused by the parameter value changes, and the app functionalities are
not impacted. Sometimes the energy fluctuation is too large and abrupt to be
filtered out by the $t$-test threshold, while in other cases the app may not
function normally with the parameter change.

\subsection{Alternative Design} \label{sec:framework-alternativeDesign}

An alternative design for estimating the energy impact of deep parameters utilizes static or dynamic analysis. For each deep parameter, we can identify the dependent code segments by performing static data and control dependence analysis~\cite{10.1145/2594291.2594299} or dynamic taint analysis~\cite{10.5555/1924943.1924971,8010886}. During execution, the energy consumption of the dependent code segments is measured and attributed to the corresponding parameter. Such an approach has the advantage of measuring the energy impacts of all deep parameters at once, but it also faces a number of challenges.

First, the dependencies between a deep parameter and the
relevant code segments are often hard to track, or
require \emph{ad hoc} customizations to achieve good
coverage. For example, we observed that many
dependencies span programming language boundaries,
or involve inter-process or inter-device (\eg client-server) communication.

Second, it is also hard to determine the right
granularity of the code segments for dependence analysis.
Dependence analysis at the branch level
of branches has the advantage that the causal relationship between parameter values and branch conditions is easy to analyze. However, we observed
that many deep parameters affect app energy consumption in ways
other than controlling branch conditions. For example,
app energy consumption can be affected by controlling the
timer duration or thread count. Alternatively, we
can perform the analysis at the method granularity by tracking
the dependency between deep parameters and method call
arguments. However, the relation between the parameter value and the
method energy consumption is often opaque, if there is any relation at all.

In view of these challenges, we chose to apply the parameter mutation approach detailed earlier and leave improvements on program analysis tools for future work.

\subsection{Implementation}

We implemented our framework in 3.5 KLoC:
  parameter analysis and mutation (in Spoon~\cite{doi:10.1002/spe.2346}); %, a Java analysis framework.
  UI automation (Appium~\cite{appium}); % the test automation framework Appium.
  and
  coverage analysis (JaCoCo~\cite{jacoco}).
  App source code is required since we perform parameter analysis by examining the source code syntax tree.

The app needs to be rebuilt for each parameter mutation. Building apps and
running test scripts are done in parallel, so that both the desktop and the
phone can be fully utilized. We also make our test framework fully reentrant,
and thus different tests can run on different phones independently.

\section{Experimental Design} \label{sec:ExperimentalDesign}

We perform our experiments using \ExperimentNApps popular open-source Android apps. We choose
the apps from 16 different categories to make sure our findings are not
restricted to certain app categories. As we use Spoon to analyze app source
code, we restrict ourselves to apps mostly (>70\%) written in Java. Table~\ref{tab:app}
summarizes each app.

\begin{table*}
\caption{Tested apps and their test configurations. The popularity statistic (installs) is from Google Play.}
%\vspace{-0.1in}
\label{tab:app}
\centering
      {\small
        \begin{tabular}{cccclcc}
\toprule
App (Abbr.) & Category & Installs & Version & Test Scenario & $t_s$ & $t_d$ \\
\midrule
SAI (\texttt{sai}) & App installer & 5M+ & 4.5 & Install 2 apps & 0.02 & 0.01 \\
ConnectBot (\texttt{cb}) & SSH client & 4M+ & 1.9.7 & View 6 Python files using vi & 0.03 & 0.03 \\
AnySoftKeyboard (\texttt{ask}) & Keyboard & 2M+ & 1.10-rc4 & Type username and password & 0.03 & 0.01 \\
KeePassDroid (\texttt{kpd}) & Password manager & 2M+ & 2.5.12 & Copy 8 password entries & 0.03 & 0.03 \\
Amaze File Manager (\texttt{amaze}) & File manager & 1M+ & 3.4.3 & Move a picture and delete a picture & 0.03 & 0.03 \\
AntennaPod (\texttt{ap}) & Podcast client & 691K+ & 1.8.3 & View 6 episode descriptions & 0.03 & 0.03 \\
OpenKeychain (\texttt{ok}) & Encryption & 538K+ & 5.7.5 & Encrypt and decrypt a file & 0.02 & 0.01 \\
Slide for Reddit (\texttt{slide}) & Online community & 222K+ & 6.3 & View 3 posts in 3 subreddits & 0.08 & 0.02 \\
Conversations (\texttt{conv}) & Instant messenger & 127K+ & 2.8.9 & Send 10 random messages & 0.04 & 0.01 \\
Download Navi (\texttt{dn}) & Download manager & 75K+ & 1.4 & Download a 100MB file & 0.03 & 0.02 \\
Wikimedia Commons (\texttt{wc}) & Image sharing & 69K+ & 2.13 & View 4 images & 0.08 & 0.02 \\
Etar Calendar (\texttt{etar}) & Calendar & 39K+ & 1.0.26 & Create 3 events & 0.08 & 0.03 \\
IPFS Lite (\texttt{ipfs}) & P2P Browser & 4K+ & 2.5.4 & View 5 Wikipedia articles & 0.06 & 0.03 \\
F-Droid (\texttt{fdroid}) & App store & N/A & 1.8 & View 3 app descriptions & 0.06 & 0.03 \\
F-Droid Build Status (\texttt{build}) & Continuous delivery & N/A & 2.8.0 & View 5 build logs & 0.03 & 0.03 \\
RadarWeather (\texttt{rw}) & Weather & N/A & 4.4 & View weather of 5 cities & 0.04 & 0.03 \\
\bottomrule
        \end{tabular}
        }
\vspace{-2mm}
\end{table*}

We design one test scenario for each app based on their typical usages
(Table~\ref{tab:app}). The lengths of the test cases are between 30 and 60
seconds.

The stability threshold $t_s$ is determined by first running the experiments
without the threshold. We then draw the CDF of the normalized standard
deviations as in Figure~\ref{fig:ref-sd-cdf}, and choose the turning point of
the CDF curve as $t_s$. To determine the other threshold $t_d$, we rerun the
parameters that pass the standard $t$-test ($t_d = 0$), and do the $t$-test again
on the new measurements. Those that only pass the first $t$-test are considered
due to energy fluctuations instead of the parameters themselves. Then we choose
the minimum $t_d$ that filters out the fluctuating ones while keeping the rest.

\subsection{Energy Measurement}

We run the experiments on two Pixel 2 phones, which are connected via USB to install app variants and accept UI automation commands.
Since power meter readings are inaccurate when devices are connected~\cite{10.1145/3106237.3106244}, power models are used to measure energy consumption.
We use well-established utilization-based power models~\cite{10.1145/2789168.2790107,10.1145/3064176.3064206} for CPU and GPU energy, and finite-state machine-based modeling~\cite{10.1145/2736277.2741635,10.1145/2307636.2307658,180294,6888881,10.1145/1999995.2000026} for WiFi.
We calculate the power of each hardware component by collecting the relevant data (state and frequency information for CPU/GPU; transmission log for WiFi) using ftrace \cite{2020ftrace} and feeding them into the power models.
As only the energy consumption of hardware components with power models can be calculated, we did not test apps that use specialized hardware components like hardware codecs or GPS.\footnote{
Prior work on GPS parameter tuning~\cite{10.1145/3236024.3236076} used long-running experiments instead, and they estimated energy consumption by reading the battery level.
} 
\section{Results and Findings} \label{sec:results}

\subsection{Parameter Extraction}

\begin{table}
	\caption{
	Effectiveness of combining coverage- and heuristic-based filtering.
	%Number of constants left without filtering, with only coverage-based filtering, with only heuristic-based filtering, and with both enabled. 
	Other apps are omitted for space.
	}
	\label{tab:filter-breakdown}
	\centering
	\small
	\begin{tabular}{c|ccc|ccc}
		\toprule
		& \multicolumn{3}{c|}{\texttt{cb}} & \multicolumn{3}{c}{\texttt{kpd}} \\
		& Num & Bool & Enum & Num & Bool & Enum \\
		\midrule
		No filtering & 12402 & 844 & 269 & 1218 & 575 & 164 \\
		Coverage & 1277 & 311 & 73 & 310 & 142 & 29 \\
		Heuristic & 9404 & 236 & 112 & 606 & 184 & 113 \\
		Combined & 451 & 115 & 31 & 131 & 71 & 13 \\
		\bottomrule
	\end{tabular}
\vspace{-2mm}
\end{table}

\begin{figure}
	\includegraphics[width=\columnwidth,trim=0 30 0 20]{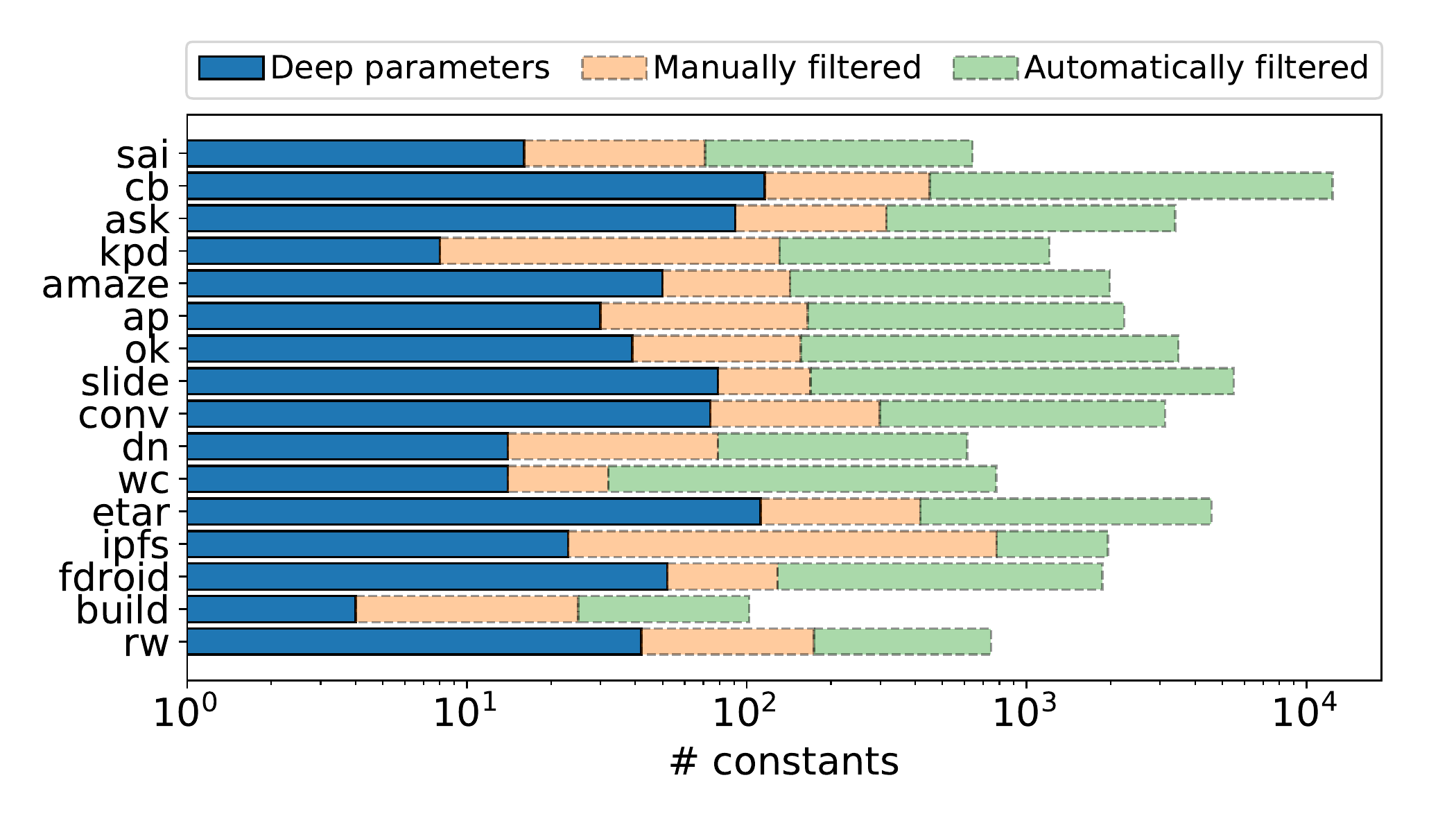}
	\caption{
	Number of numeric parameters identified (blue), and of constants filtered out by manual (beige) after automatic (green) filtering.
	Other parameter types are omitted for space.
	}
	\label{fig:filter-bar}
\end{figure}

To speed up testing, we filter out unused and non-parameter constants by combining both automatic and manual filtering.
\cref{fig:filter-bar} shows the effect of each filtering method for numeric constants.
%number of constants filtered out by each filtering method, as well as the number of parameters left after filtering. Due to the page limit, we only show
%the breakdown for numeric parameters.
Automatic filtering filters out 92.1\% of the numeric constants.
Manual filtering filters out another 6.2\%, which further speeds up testing and leaves us on average 48 deep parameters per app.
For Boolean constants the proportions are 90.3\% automatic and 5.5\% manual, leaving on average 40 deep parameters per app.
For enumerator references, the proportions are 88.9\% and 9.2\%	, and on average 15 are left.

To measure the effectiveness of each automatic filtering method, we turn them
on and off individually and calculate the number of constants filtered out.
Per~\cref{tab:filter-breakdown}, coverage-based filtering alone filters out on average 86.2\% constants, while heuristic-based filtering alone filters out 31.1\%.
Combining them further improves the filtering efficiency to 94.8\%, reducing subsequent manual filtering effort.

\begin{figure}
\centering
\subcaptionbox{The energy drain of \texttt{slide} with respect to the post fetching batch size under two usage scenarios.\label{fig:slide}}
{\includegraphics[width=.48\columnwidth]{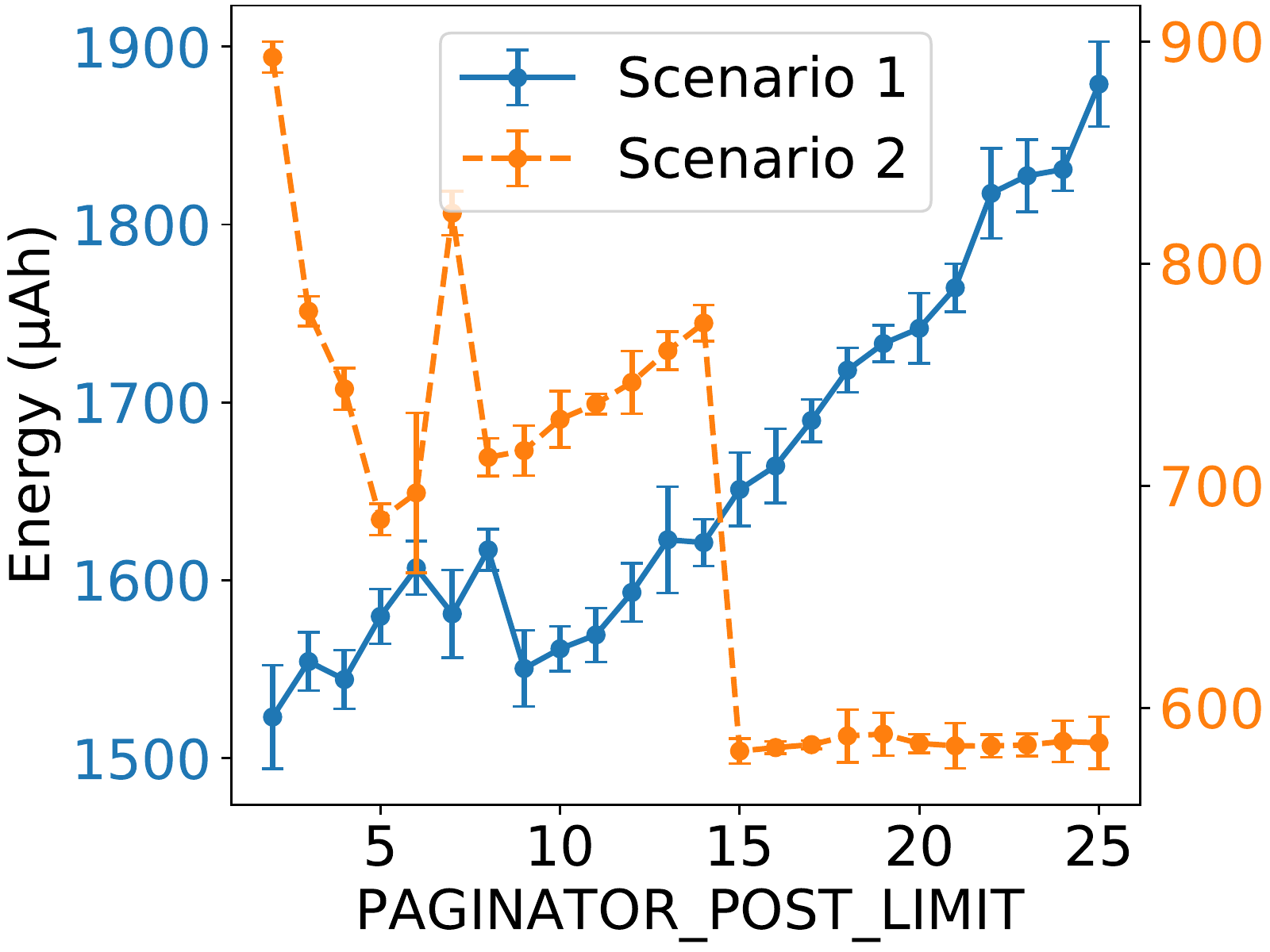}}
\ 
\subcaptionbox{The energy drain of \texttt{cb} under different channel buffer sizes. ``$\times$'' corresponds to the original (energy efficient) parameter value.\label{fig:cb-buffer}}
{\includegraphics[width=.48\columnwidth]{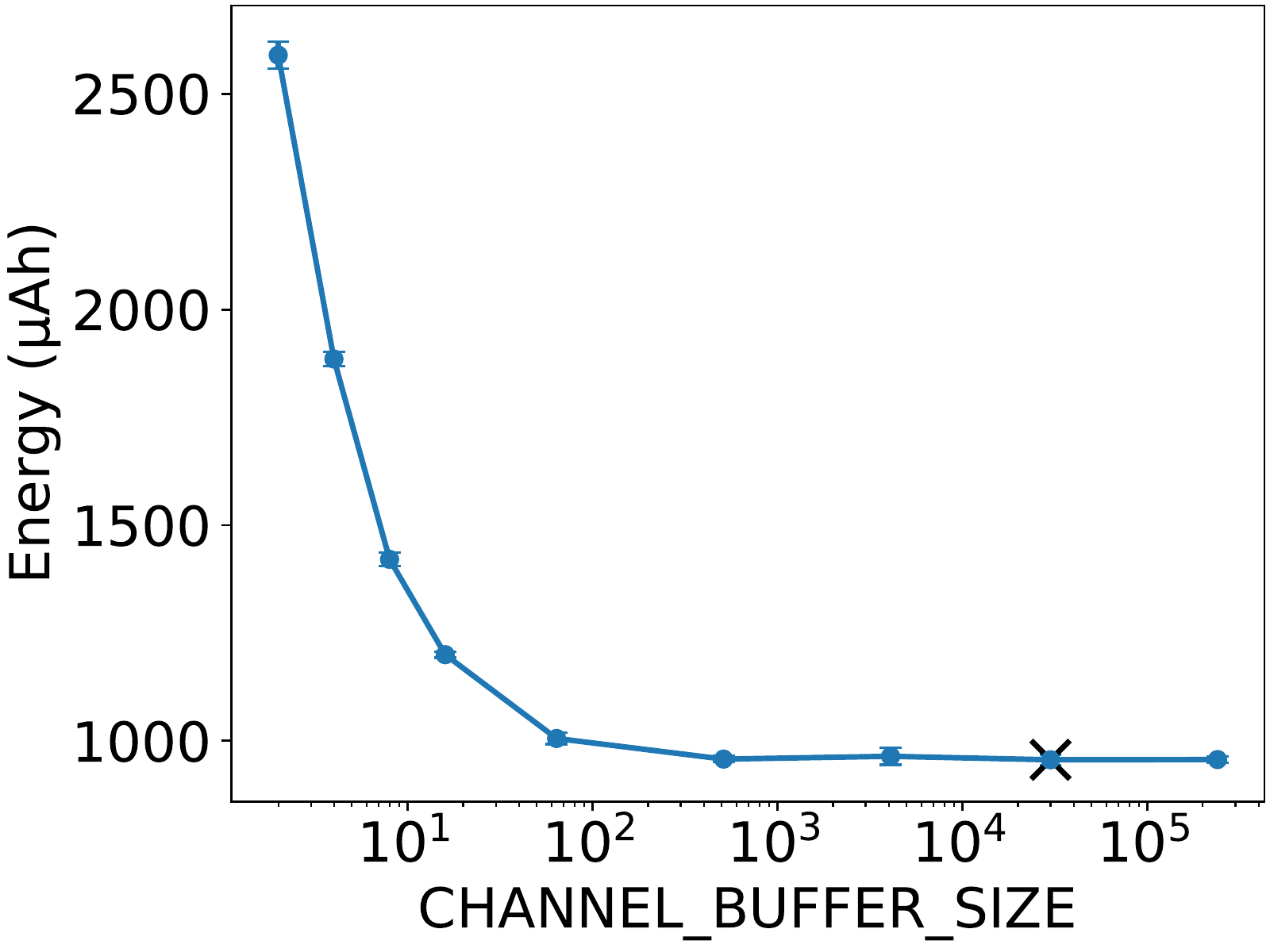}}
\caption{Energy effects of two deep parameters. Error bars represent the standard deviation of 5 runs. Note y-axes do not start from 0.}
\vspace{-4mm}
\end{figure}

\subsection{RQ3: Parameter-Induced Energy Inefficiency}

To see whether energy-reducing deep parameters are common among apps, we mutate each
deep parameter and measure the energy drain of each variant. After filtering the
constants in the 16 apps, we get in total 764 numeric parameters, 647 Boolean
parameters, and 233 enumerator parameters. In testing all the parameters, we
run the automated tests 15040 times (3008 parameter values with 5 runs each),
which is 596 hours (25 days) of phone execution time.

Table~\ref{tab:result} shows a summary of the test results.
Out of the 1644 deep parameters tested, we observed reduced energy drain for 28
parameters. We then manually examined the 28 parameters, and found that only
2 numeric parameters really reduce the energy drain of the app without
breaking app functionality.

\begin{tcolorbox}
    \textbf{Finding 3}:
	Parameter-induced energy inefficiency is uncommon among apps. Out of
	the 1644 deep parameters from the 16 apps, only 2 reduce energy drain without breaking app functionality.
\end{tcolorbox}

\begin{table}
\caption{
    Number of deep Parameters (P) tested for each app and parameter type,
	number of parameters that appear to Reduce the energy drain (R) during the test,
	and
	number of parameters manually Validated (V) to be energy-reducing.
	}
\label{tab:result}
\centering
      {\small
\begin{spreadtab}{{tabular}{>{\ttfamily}c|ccc|ccc|ccc}}
\toprule
& @ \multicolumn{3}{c|}{Numeric} & @ \multicolumn{3}{c|}{Boolean} & @ \multicolumn{3}{c}{Enum} \\
& @ P & @ R & @ V & @ P & @ R & @ V & @ P & @ R & @ V \\
\midrule
@ sai & 16 & 2 & 0 & 22 & 0 & 0 & 10 & 0 & 0 \\
% UI disappear, UI disappear
@ cb & 116 & 2 & 0 & 17 & 0 & 0 & 10 & 0 & 0 \\
% ref is high, hight inconsistent with buffer size, non-tunable
@ ask & 91 & 0 & 0 & 35 & 0 & 0 & 17 & 0 & 0 \\
@ kpd & 8 & 0 & 0 & 37 & 0 & 0 & 5 & 0 & 0 \\
@ amaze & 50 & 0 & 0 & 25 & 0 & 0 & 15 & 0 & 0 \\
% non-tunable
@ ap & 30 & 0 & 0 & 33 & 2 & 0 & 18 & 0 & 0\\
% no cover image, mut is low
@ ok & 39 & 0 & 0 & 55 & 4 & 0 & 14 & 0 & 0 \\
% UI disappear, UI disappear, cannot decrypt, no compression
@ slide & 79 & 2 & 1 & 168 & 0 & 0 & 29 & 0 & 0 \\
% ref is high, PAGINATOR_POST_LIMIT, non-tunable
@ conv & 74 & 3 & 0 & 58 & 1 & 0 & 24 & 0 & 0 \\
% script delay, UI delay, mut is low, ref is high
@ dn & 14 & 1 & 0 & 50 & 0 & 0 & 9 & 0 & 0 \\
% UI delay, non-tunable
@ wc & 14 & 1 & 0 & 1 & 1 & 0 & 7 & 0 & 0 \\
% ref is high, non-reproducible
@ etar & 112 & 1 & 0 & 65 & 0 & 0 & 39 & 0 & 0 \\
% non-reproducible
@ ipfs & 23 & 1 & 1 & 18 & 2 & 0 & 3 & 0 & 0 \\
% low p2p ping frequency, broken p2p connection, no js support
@ fdroid & 52 & 1 & 0 & 32 & 0 & 0 & 10 & 0 & 0 \\
% ref is high, non-tunable, non-tunable
@ build & 4 & 1 & 0 & 10 & 0 & 0 & 15 & 0 & 0 \\
% ref is high
@ rw & 42 & 1 & 0 & 21 & 0 & 0 & 8 & 2 & 0 \\
% api limit, ref is high, UI becomes vertical, non-tunable
\midrule
@ \textnormal{Total} & sum(b3:b18) & sum(c3:c18) & sum(d3:d18) & sum(e3:e18) & sum(f3:f18) & sum(g3:g18) & sum(h3:h18) & sum(i3:i18) & sum(j3:j18) \\
\bottomrule
\end{spreadtab}
      }
\vspace{-2mm}
\end{table}

\subsubsection{The True Positives}

One energy-reducing parameter is identified in the Reddit client \texttt{slide}.
While browsing the posts in a subreddit, the app fetches posts from the server.
A batch of posts will be fetched each time, and each post will be processed immediately after being fetched.
The energy used for processing is wasted if posts are fetched but not displayed.

The optimal batch size depends on the usage scenario. In \cref{fig:slide}, we measured the energy consumption with respect to the batch size for two different scenarios: view 3 posts in 3 subreddits (Scenario 1) and scroll 5 times in a subreddit feed (Scenario 2). The first scenario favors a smaller batch size, as it only loads the first screen for each subreddit. The second scenario favors a larger batch size as it loads more posts.

The other energy-reducing parameter, identified in the P2P browser \texttt{ipfs}, controls the ping interval to multiple peers.
By reducing the ping frequency from once every second to once every 8 seconds, the app energy usage is reduced by 12.1\% due to less frequent WiFi usage.

\subsubsection{The False Positives}

The other deep 26 parameters appear to reduce energy drain for three different
reasons.
The 3 numeric parameters in \texttt{conv} happened to correspond to the three reasons.
Thus, we use these parameters to illustrate. %explain the reasons using the 3 numeric parameters as examples.

\begin{figure}
\includegraphics[width=1.1\columnwidth,trim=0 5 0 20]{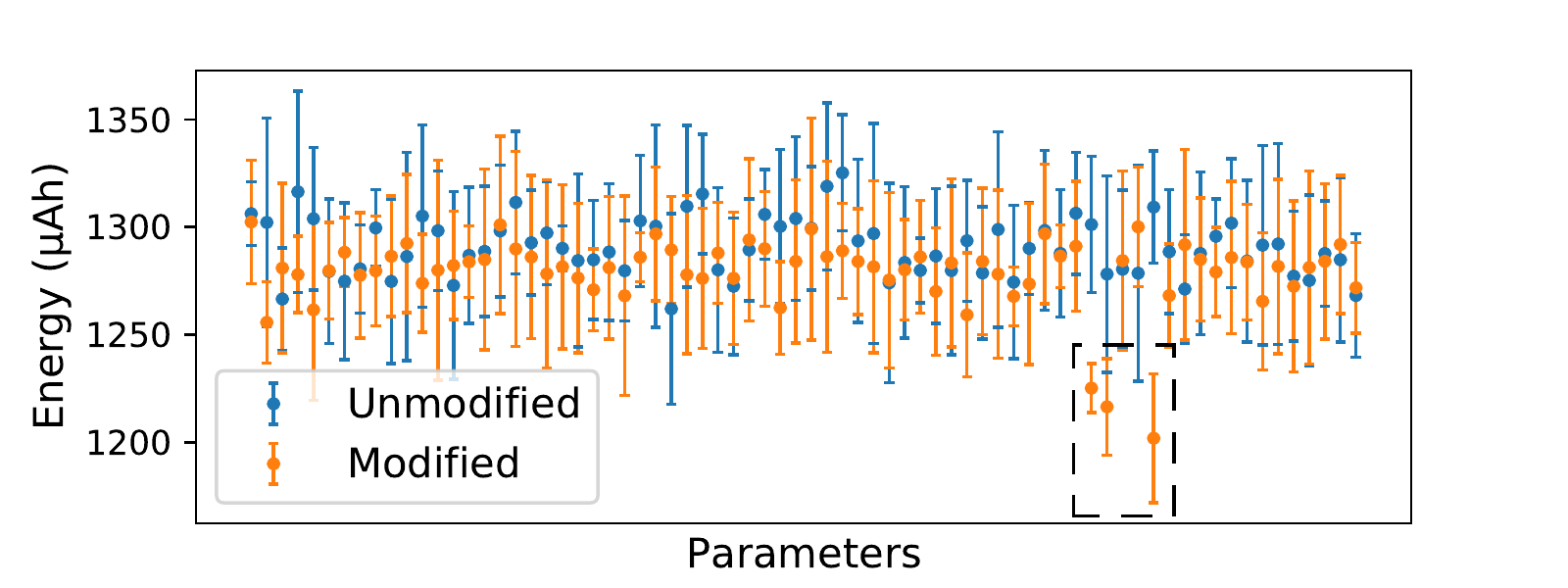}
\vspace{-0.2in}
\caption{
    For each of the 74 numeric parameters in \texttt{conv}, the energy consumption of the parameter value consuming the least energy (Modified) vs. the original parameter value (Unmodified).
	Error bars represent the standard deviation of 5 runs.
	Data points in dashed box have statistically significant reduction in energy drain.}
\label{fig:conv}
\vspace{-4mm}
\end{figure}

Figure~\ref{fig:conv} shows the test results of the 73 numeric parameters in
\texttt{conv}.
All standard deviations are within 4\% ($t_s$) of the corresponding means. Most parameter values have energy
consumption very close to the original values. The only three data
points that have statistically significant energy difference are those
in the dashed box (5.8\%, 4.8\%, 8.2\%).

The rightmost data point represents the most common reason (12 of the 26
parameters): Even though we try to reduce false positives caused by the
stochastic energy drain through measures like back-to-back testing, hypothesis
testing, and stability threshold, sometimes the changes are too large and
abrupt that the framework treats them as real energy reductions. The way to
identify such cases is to rerun the tests and see if the energy reduction can
be reproduced.

The data point in the middle corresponds to the second reason (13 of the 26
parameters): energy reduction at the cost of broken or degraded functionality.
The \texttt{conv} parameter controls the refresh rate of various UI elements.
By increasing the refresh interval from 500ms to 4000ms, the app energy consumption is reduced by 4.8\%.
However, when sending a text message, the message will take 4 seconds to appear in the conversation view. % only several seconds after it has disappeared in the text editing box. %, resulting in a degraded user experience.
Other parameters have problems like blanking out the app screen or disappearing all images.

The last and only one in its category is due to unexpected interaction between the app and our test automation script.
When measuring energy consumption, we omit the initialization phase of the app and only measure the target test scenario.
As we do not know exactly when all initializations
are done, there may be some lingering initialization tasks after we have
entered the test scenario, and their energy consumption will also be measured.
This is not a problem as long as we enter the test scenario after the same
delay. But the \texttt{conv} parameter delays entering the test scenario until
all initializations have been done, leading to the apparent energy reduction.

\subsection{RQ4: Parameters' Energy Effect}

In the last section, we showed that deep parameter-induced energy inefficiency is
uncommon among apps, and discussed the 2 energy-reducing parameters we
discovered. In this section, we consider the ``Why'' question:
  \emph{Why do deep parameters commonly not affect energy usage?}

To answer the question, we manually examine the 143 deep parameters in \texttt{cb},
and try to figure out their energy effects by understanding their semantics in
the context of the source code and testing additional parameter values. We
finally classified them into 3 categories based on their energy effect: having
no energy effect, having limited energy effect, and having energy effect under
extreme values.

\subsubsection{Deep Parameters with No Energy Effect}

71 of the 143 parameters fall in this group and are further divided into two
representative types. The first type of such parameters has binary effects.
When the parameter value is in the valid range, the app works the same way
regardless of the exact parameter value. On the other hand, the app breaks if
the parameter value is in the invalid range. Lines 2-3 in~\cref{fig:param} shows two examples.
Line 2 creates a buffer for version string parsing.
App behavior is preserved when the buffer is big enough to hold the version string, but incorrect parsing occurs when the buffer is too small.
In the second example, if the timeout of the socket is
longer than the server's response latency, the socket communication works
normally regardless of the exact timeout value (0 means indefinite timeout);
the connection breaks if the timeout value is too small.

The other type of deep parameters without energy effect is due to limitations of our energy measurement methodology.
We use power models to calculate the energy consumption of each hardware component.
The change in energy consumption of a hardware component cannot be captured if the corresponding power model is missing.
For example, the choice of colors can affect the energy consumption of OLED displays~\cite{10.1145/2568225.2568321}, but we omitted it since measuring the OLED energy consumption is expensive (we would have to record every frame).

\subsubsection{Deep Parameters with Limited Energy Effects}

61 deep parameters fall in this category. Each parameter is typically attached to a
certain component of the app. Thus, the energy effect of the parameter depends
on both the total energy consumption of the component and the importance of the
parameter in the component. The logging component typically consumes a limited
amount of energy for most apps; the parameters controlling logging levels will
thus have limited energy effect. On the other hand, although UI is energy
expensive, the energy consumption of UI rendering mainly depends on the
structure of the UI element tree, instead of the precise positions and sizes of
the individual UI elements, and thus these UI parameters also have a limited energy
effects.

\subsubsection{Deep Parameters Having Effects under Extreme Values}
\label{subsec:extreme}

To see how energy drain can be affected by extreme parameter values, we will
first look at an example. Figure~\ref{fig:cb-buffer} shows the energy drain of
\texttt{cb} under varying \texttt{CHANNEL\_BUFFER\_SIZE}, which controls the
size of the \texttt{stdout} and \texttt{stderr} buffers attached to the
terminal. Extremely small buffers divide terminal outputs into small chunks,
and processing them one by one adds overhead. Such extreme values only occupy a
tiny fraction of the valid value range, thus are not captured by our framework.
However, a developer is also unlikely to pick such extreme values if she
understood the meaning of the parameter.

Apart from buffer sizes, making the font size of the terminal extremely small
also increases energy consumption drastically. In total 9 parameters in
\texttt{cb} are of this kind. Such parameters also exist in other apps. For
example, the UI update frequency parameter discussed in the last section also
only exhibits an energy effect when the update interval is extremely long.
Similarly, a developer will not choose such extreme values if she considered
the semantics of the parameters. Basically, developers can typically avoid such
extreme values based on their domain knowledge.

\begin{tcolorbox}
    \textbf{Finding 4}:
	Most deep parameters either have no energy effect, limited energy effect, or
	only have energy effect under extreme values. We expect developers would typically avoid
	such extreme values based on their domain knowledge. 
\end{tcolorbox}

\section{Discussion and Future Work}

\paragraph{Potential impact factors to energy consumption}
Our work is the first systematic attempt to understand the energy impact of deep parameters in mobile apps.
\emph{Across \ExperimentNApps apps, we found that mobile deep parameters did not have a significant impact on app energy.} We conjecture three possible explanations.
First, it may be that the app's design --- the software architecture and design patterns~\cite{martin2000design} --- has a dominant effect on the app's energy usage~\cite{7884614,cruz2019catalog,9054858}.
Second, it may be that our constraint was too strong; mobile apps may have to sacrifice user experience or remove features to conserve energy.
Third, while individual parameters cannot move the needle, tuning them in combination might have a bigger impact~\cite{10.1145/2528265.2528267,kolesnikov_relation_2019}.
Each of these possibilities is a direction for further study.

\paragraph{Automatic support for tuning deep parameters}
Most parameter tuning systems work only with the ``formal'' parameters exposed by developers in a central repository (cf. \cref{sec:related}).
This design assumes that developers have identified and centralized their parameters.
However, when exposing parameters by hand, it is difficult to anticipate the needs of future use cases.
Wang \etal~\cite{10.1145/3173162.3173206} discussed difficulties in tuning database systems because developers had hard-coded deep parameters instead of exposing them for tuning.

One strength of our deep parameter-identification framework (\cref{fig:workflow}) is that we \emph{automatically} identify deep parameters.
In the future, we plan to apply our deep parameter search framework to other classes of software (\eg database systems) and help discover those important deep parameters.
In these contexts, we will develop a unified parameter tuning approach that merges formal and deep parameters.

\iffalse
To conduct performance tuning for software that has manually exposed configuration parameters, one can simply play with the set of exposed parameters. However, the process of deciding which
parameter to expose is ad-hoc and completely depends on the developers'
experience.  The issues show that exposing parameters by hand cannot fulfill the
needs of all usage cases, even if hundreds of parameters have already been
exposed.
\fi

\paragraph{Large-scale energy measurement}
Current mobile phone energy measurement methods (both power monitors and power models) rely on real phones, making energy measurement unscalable.
In our experiments, it took on average \ExperimentAvgTime minutes to test each parameter, which means roughly 1.5 days per app.
Accurate energy measurement in virtualized environments will enable larger-scale experiments on energy optimization.
Accurate emulation of the hardware states and frequencies for power modeling is one possible direction.

\section{Threats to Validity} \label{sec:valid}

\paragraph{Internal Validity} % Causality: the strength of assigning causes to outcomes

There are several threats to internal validity.
\ul{Survey}: Although we refined our survey instrument through pilot studies, it has not been validated~\cite{kitchenham2008personal}. We assume our respondents replied honestly.
\ul{Energy experiments}: Energy changes might be due to factors other than the mutated parameter.
Such factors include changes in timing, network conditions, and external service behaviors.
This threat is mitigated by using automated testing and repeated trials for each deep parameter.
%This is mitigated by the fact that we only received responses from direct contacts that are known to be Android developers, and online communities of Android developers.

\paragraph{External Validity} % Generalizability: validity of applying your conclusions outside the study context

% Generalize to different parameters.
\iffalse
Our findings may not generalize to different kinds of parameters.
Our experiments only consider app parameters located in the app's source code, without looking at those stored in other places --- \eg Android resource files or local databases (\cref{fig:param-location}).
For Android resource files, this threat is mitigated by the fact that such parameters can be declared in either source files or Android resource files (lines 4-5 in \cref{fig:param}), so that we can still analyze those declared in the source code.
\fi
%Our experiments do not consider parameters outside the control of the app developer, \eg in 3rd-party libraries or the Android framework.
%Although tuning these parameters can reduce app energy usage~\cite{10.1145/3067695.3082519,10.1145/3286978.3287014,naik2014categorizing}, they are not available to the app developers themselves.

% Generalize to different apps.
Our findings may not generalize to different classes of software~\cite{nagappan2013diversity}.
Within Android apps, there may be differences between the \ExperimentNApps open-source apps we investigated vs. (1) commercial apps, and (2) apps that are deliberately designed to be energy-efficient.
For some insight on this threat, most of our survey respondents develop Android apps commercially.
They indicated that their parameters and parameterizations were not designed for energy efficiency.
%Nevertheless, our testing framework is available publicly to enable generalizability tests. %testing others can validate on commercial apps.

\paragraph{Construct Validity} % Are constructs (concepts) clearly defined? Are they measured from multiple perspectives?

We define a deep parameter as a constant, and our experimental design preserves each parameterization throughout the lifetime of each measurement.
Similarly, we tune deep parameters on a (static) per-class basis rather than a (dynamic) per-instance basis. %, losing the flexibility that parameters in different class instances can be fine-tuned independently.
This definition is generally consistent with how these constructs are defined in the apps we studied.
However, %it is possible that different results would emerge if parameter values were varied dynamically during a measurement.
  a dynamic notion of deep parameters might affect our results; for example, the energy-optimal parameter choice has been shown to vary dynamically for GPS parameters in Android apps~\cite{10.1145/3236024.3236076}.

\section{Related Work}
\label{sec:related}

\paragraph{Configuration tuning}

Tuning software configuration parameters for better performance is a common practice for many classes of software.
Recently, many automated configuration tuning systems are proposed, either for arbitrary configurable systems~\cite{10.1145/3127479.3128605,10.1145/3173162.3173206,10.1145/3379469,8952456,10.1145/2739480.2754648,10.1145/2714064.2660196}, or for specific application types~\cite{10.1145/3035918.3064029,246156}.
Tuning is conducted either
  offline, optimizing parameter values for a fixed workload and environment~\cite{215953,10.1145/3342195.3387526},
  or
  online, dynamically reconfiguring the target system to adapt to changes~\cite{234950,10.1145/3126908.3126951}.
%Many of the automatic tuners are powered by machine learning.

\iffalse
The major difference between our work and all the systems above, is that they all rely on configuration parameters that are explicitly exposed by the software developers.
For example, many database systems store all configuration parameters in a single table~\cite{10.1145/3299869.3300085}.
In contrast, guided by development practices in the Android community (\cref{sec:DevSurvey}), we extract deep parameters that scattered around the app source code. % which are often overlooked even by the developers.
\fi
Most such works are focused on systems software, \eg file systems and databases.
Only Bokhari \etal~\cite{10.1145/3067695.3082519,10.1145/3286978.3287014} and Canino \etal~\cite{10.1145/3236024.3236076} have considered deep parameters in Android apps.
They focus on deep parameters in specific app components.
Our work is the first step to understanding the energy impact of deep parameters in general Android apps; we study the energy impact of single parameters, and leave combinatorial tuning to future work.

\paragraph{Performance modeling}

Many works~\cite{10.1145/2786805.2786845,10.1145/3236024.3236074,8812049,10.1145/3342195.3387554,velez_configcrusher_2020} focus on building a performance model for a certain application and workload. A performance model is a mathematical function where the domain is the configuration parameters and the codomain is the performance. Performance optimization or other tasks can be further performed based on the performance model. These systems mostly rely on sampling, and they generate a better performance model by sampling efficiently. These works rely on explicitly exposed parameters.
We consider instead a program's deep parameters.

\paragraph{Energy impacts of design patterns and refactoring}

Researchers have studied the energy impacts of design patterns across software domains.
Sahin \etal~\cite{6224257} compared the power profile of data center software using design patterns against those not using.
Pinto \etal~\cite{10.1145/2714064.2660235} focus on the energy consumption of Java thread management constructs.

Refactoring energy-greedy code patterns can also reduce app energy drain.
Carette \etal~\cite{7884614} design a framework that automatically refactors Android code smells and observe reduced energy drain after correcting them.
Cruz and Abreu~\cite{cruz2019catalog} study refactorings for energy efficiency in the wild by mining source code commits, issues, and pull requests.
Couto \etal~\cite{9054858} further study the impacts of refactoring on energy consumption by applying combinations of refactorings to a large set of Android apps.
Their guidelines help developers reduce energy drain through refactoring.
Our work considers energy improvement through parameterization instead of refactoring.

\paragraph{Mobile app energy testing}

Discovering energy inefficiency through testing is an ongoing research topic.
Ding and Hu~\cite{10.1145/3064176.3064206} uncover the potential energy inefficiency during the rendering process.
Jindal and Hu~\cite{222559} discover energy-inefficient components by comparing them with other apps with similar functionalities.
Jabbarvand \etal~\cite{8812097} enhances UI automated testing techniques to cover energy-heavy APIs, but lack proper oracles for unknown energy defects.
As a first step towards automated energy test oracle construction, their subsequent work~\cite{10.1145/3368089.3409677} employed deep learning to determine energy efficiency based on lifecycle and hardware states.
Li \etal~\cite{10.1145/3395363.3397350} classified mobile app energy issues into 6 categories and proposed different methods to detect the energy issues of each category.
Our work focuses on constructs at a finer granularity, and is complementary to those works.

\section{Conclusion}

We studied the energy impact of mobile deep parameters.
We used a developer survey to understand the perceived impact, and a systematic experiment to understand the actual impact.
Our survey showed that many app developers are uncertain about and ignore the energy impact of deep parameters.
%We designed a parameter-centric energy profiling framework to systematically measure the energy impact of the parameters.
Our experiment and analysis with \ExperimentNApps apps showed that single-parameter-induced energy inefficiency is uncommon.
However, in order to more fully explore energy-feature tradeoffs, in future work we plan to explore energy optimization opportunities through tuning combinations of deep parameters as well as through non-functionality-preserving parameterizations.
For now, it appears that mobile app developers can ignore the energy impact when choosing deep parameter values --- they will not substantially degrade their app's energy performance.

\raggedbottom

\section*{Acknowledgments}
This work was supported in part by NSF CSR-1718854.

\newpage
\balance

\bibliographystyle{IEEEtran}
\bibliography{IEEEabrv,main}

\begin{thebibliography}{10}
\providecommand{\url}[1]{#1}
\csname url@samestyle\endcsname
\providecommand{\newblock}{\relax}
\providecommand{\bibinfo}[2]{#2}
\providecommand{\BIBentrySTDinterwordspacing}{\spaceskip=0pt\relax}
\providecommand{\BIBentryALTinterwordstretchfactor}{4}
\providecommand{\BIBentryALTinterwordspacing}{\spaceskip=\fontdimen2\font plus
\BIBentryALTinterwordstretchfactor\fontdimen3\font minus
  \fontdimen4\font\relax}
\providecommand{\BIBforeignlanguage}[2]{{%
\expandafter\ifx\csname l@#1\endcsname\relax
\typeout{** WARNING: IEEEtran.bst: No hyphenation pattern has been}%
\typeout{** loaded for the language `#1'. Using the pattern for}%
\typeout{** the default language instead.}%
\else
\language=\csname l@#1\endcsname
\fi
#2}}
\providecommand{\BIBdecl}{\relax}
\BIBdecl

\bibitem{6682059}
C.~Wilke, S.~Richly, S.~Götz, C.~Piechnick, and U.~Aßmann, ``Energy
  consumption and efficiency in mobile applications: A user feedback study,''
  in \emph{2013 IEEE International Conference on Green Computing and
  Communications and IEEE Internet of Things and IEEE Cyber, Physical and
  Social Computing}, 2013, pp. 134--141.

\bibitem{MobilePhonesMatterDevelopingWorld}
A.~K. Pramanik, ``{The Technology That's Making a Difference in the Developing
  World},''
  \url{https://www.usglc.org/blog/the-technology-thats-making-a-difference-in-the-developing-world/},
  2017, accessed: 2021-04-20.

\bibitem{MobilePhoneBenefits}
P.~Research, ``{Mobile Connectivity in Emerging Economies},''
  \url{https://www.pewresearch.org/internet/2019/03/07/mobile-connectivity-in-emerging-economies/},
  accessed: 2021-04-20.

\bibitem{androidbattery}
\BIBentryALTinterwordspacing
(2021, Feb.) Optimize for battery life. [Online]. Available:
  \url{https://developer.android.com/topic/performance/power}
\BIBentrySTDinterwordspacing

\bibitem{iosenergy}
\BIBentryALTinterwordspacing
Energy efficiency and the user experience. [Online]. Available:
  \url{https://developer.apple.com/library/archive/documentation/Performance/Conceptual/EnergyGuide-iOS/index.html}
\BIBentrySTDinterwordspacing

\bibitem{10.1145/3127479.3128605}
Y.~Zhu, J.~Liu, M.~Guo, Y.~Bao, W.~Ma, Z.~Liu, K.~Song, and Y.~Yang,
  ``Bestconfig: Tapping the performance potential of systems via automatic
  configuration tuning,'' in \emph{Proceedings of the 2017 Symposium on Cloud
  Computing}, ser. SoCC ’17.\hskip 1em plus 0.5em minus 0.4em\relax New York,
  NY, USA: Association for Computing Machinery, 2017, p. 338–350.

\bibitem{10.1145/3035918.3064029}
D.~Van~Aken, A.~Pavlo, G.~J. Gordon, and B.~Zhang, ``Automatic database
  management system tuning through large-scale machine learning,'' in
  \emph{Proceedings of the 2017 ACM International Conference on Management of
  Data}, ser. SIGMOD ’17.\hskip 1em plus 0.5em minus 0.4em\relax New York,
  NY, USA: Association for Computing Machinery, 2017, p. 1009–1024.

\bibitem{10.14778/3352063.3352129}
G.~Li, X.~Zhou, S.~Li, and B.~Gao, ``Qtune: A query-aware database tuning
  system with deep reinforcement learning,'' \emph{Proc. VLDB Endow.}, vol.~12,
  no.~12, p. 2118–2130, Aug. 2019.

\bibitem{10.1145/3236024.3236076}
A.~Canino, Y.~D. Liu, and H.~Masuhara, ``Stochastic energy optimization for
  mobile gps applications,'' in \emph{Proceedings of the 2018 26th ACM Joint
  Meeting on European Software Engineering Conference and Symposium on the
  Foundations of Software Engineering}, ser. ESEC/FSE 2018.\hskip 1em plus
  0.5em minus 0.4em\relax New York, NY, USA: Association for Computing
  Machinery, 2018, p. 703–713.

\bibitem{10.1145/3067695.3082519}
M.~A. Bokhari, B.~R. Bruce, B.~Alexander, and M.~Wagner, ``Deep parameter
  optimisation on android smartphones for energy minimisation: A tale of woe
  and a proof-of-concept,'' in \emph{Proceedings of the Genetic and
  Evolutionary Computation Conference Companion}, ser. GECCO '17.\hskip 1em
  plus 0.5em minus 0.4em\relax New York, NY, USA: Association for Computing
  Machinery, 2017, p. 1501–1508.

\bibitem{10.1145/3286978.3287014}
M.~A. Bokhari, B.~Alexander, and M.~Wagner, ``In-vivo and offline optimisation
  of energy use in the presence of small energy signals: A case study on a
  popular android library,'' in \emph{Proceedings of the 15th EAI International
  Conference on Mobile and Ubiquitous Systems: Computing, Networking and
  Services}, ser. MobiQuitous '18.\hskip 1em plus 0.5em minus 0.4em\relax New
  York, NY, USA: Association for Computing Machinery, 2018, p. 207–215.

\bibitem{johnson2004mixed}
R.~B. Johnson and A.~J. Onwuegbuzie, ``Mixed methods research: A research
  paradigm whose time has come,'' \emph{Educational researcher}, vol.~33,
  no.~7, pp. 14--26, 2004.

\bibitem{x265cli}
\BIBentryALTinterwordspacing
Command line options. [Online]. Available:
  \url{https://x265.readthedocs.io/en/master/cli.html}
\BIBentrySTDinterwordspacing

\bibitem{x265api}
\BIBentryALTinterwordspacing
Application programming interface. [Online]. Available:
  \url{https://x265.readthedocs.io/en/master/api.html}
\BIBentrySTDinterwordspacing

\bibitem{10.1145/1005686.1005739}
D.~G. Sullivan, M.~I. Seltzer, and A.~Pfeffer, ``Using probabilistic reasoning
  to automate software tuning,'' in \emph{Proceedings of the Joint
  International Conference on Measurement and Modeling of Computer Systems},
  ser. SIGMETRICS '04/Performance '04.\hskip 1em plus 0.5em minus 0.4em\relax
  New York, NY, USA: Association for Computing Machinery, 2004, p. 404–405.

\bibitem{pew}
\BIBentryALTinterwordspacing
Questionnaire design. [Online]. Available:
  \url{https://www.pewresearch.org/methods/u-s-survey-research/questionnaire-design/}
\BIBentrySTDinterwordspacing

\bibitem{questionpro}
\BIBentryALTinterwordspacing
Survey design: How to design a survey that people will love to answer?
  [Online]. Available:
  \url{https://www.questionpro.com/features/survey-design/}
\BIBentrySTDinterwordspacing

\bibitem{googleplay}
\BIBentryALTinterwordspacing
Google play. [Online]. Available: \url{https://play.google.com/store/apps}
\BIBentrySTDinterwordspacing

\bibitem{10.1145/2950290.2950297}
Y.~Liu, C.~Xu, S.-C. Cheung, and V.~Terragni, ``Understanding and detecting
  wake lock misuses for android applications,'' in \emph{Proceedings of the
  2016 24th ACM SIGSOFT International Symposium on Foundations of Software
  Engineering}, ser. FSE 2016.\hskip 1em plus 0.5em minus 0.4em\relax New York,
  NY, USA: Association for Computing Machinery, 2016, p. 396–409.

\bibitem{10.1145/2635868.2635901}
M.~Allamanis and C.~Sutton, ``Mining idioms from source code,'' in
  \emph{Proceedings of the 22nd ACM SIGSOFT International Symposium on
  Foundations of Software Engineering}, ser. FSE 2014.\hskip 1em plus 0.5em
  minus 0.4em\relax New York, NY, USA: Association for Computing Machinery,
  2014, p. 472–483.

\bibitem{8952363}
D.~{Lai} and J.~{Rubin}, ``Goal-driven exploration for android applications,''
  in \emph{2019 34th IEEE/ACM International Conference on Automated Software
  Engineering (ASE)}, 2019, pp. 115--127.

\bibitem{9152712}
Y.~{He}, L.~{Zhang}, Z.~{Yang}, Y.~{Cao}, K.~{Lian}, S.~{Li}, W.~{Yang},
  Z.~{Zhang}, M.~{Yang}, Y.~{Zhang}, and H.~{Duan}, ``Textexerciser:
  Feedback-driven text input exercising for android applications,'' in
  \emph{2020 IEEE Symposium on Security and Privacy (SP)}, 2020, pp.
  1071--1087.

\bibitem{10.1145/3377811.3380347}
J.~Yan, H.~Liu, L.~Pan, J.~Yan, J.~Zhang, and B.~Liang, ``Multiple-entry
  testing of android applications by constructing activity launching
  contexts,'' in \emph{Proceedings of the ACM/IEEE 42nd International
  Conference on Software Engineering}, ser. ICSE '20.\hskip 1em plus 0.5em
  minus 0.4em\relax New York, NY, USA: Association for Computing Machinery,
  2020, p. 457–468.

\bibitem{10.1145/3377811.3380402}
Z.~Dong, M.~B\"{o}hme, L.~Cojocaru, and A.~Roychoudhury, ``Time-travel testing
  of android apps,'' in \emph{Proceedings of the ACM/IEEE 42nd International
  Conference on Software Engineering}, ser. ICSE '20.\hskip 1em plus 0.5em
  minus 0.4em\relax New York, NY, USA: Association for Computing Machinery,
  2020, p. 481–492.

\bibitem{10.1145/3377811.3380382}
J.~Wang, Y.~Jiang, C.~Xu, C.~Cao, X.~Ma, and J.~Lu, ``Combodroid: Generating
  high-quality test inputs for android apps via use case combinations,'' in
  \emph{Proceedings of the ACM/IEEE 42nd International Conference on Software
  Engineering}, ser. ICSE '20.\hskip 1em plus 0.5em minus 0.4em\relax New York,
  NY, USA: Association for Computing Machinery, 2020, p. 469–480.

\bibitem{10.1145/3395363.3397354}
M.~Pan, A.~Huang, G.~Wang, T.~Zhang, and X.~Li, ``Reinforcement learning based
  curiosity-driven testing of android applications,'' in \emph{Proceedings of
  the 29th ACM SIGSOFT International Symposium on Software Testing and
  Analysis}, ser. ISSTA 2020.\hskip 1em plus 0.5em minus 0.4em\relax New York,
  NY, USA: Association for Computing Machinery, 2020, p. 153–164.

\bibitem{10.1145/2594291.2594299}
S.~Arzt, S.~Rasthofer, C.~Fritz, E.~Bodden, A.~Bartel, J.~Klein, Y.~Le~Traon,
  D.~Octeau, and P.~McDaniel, ``Flowdroid: Precise context, flow, field,
  object-sensitive and lifecycle-aware taint analysis for android apps,'' in
  \emph{Proceedings of the 35th ACM SIGPLAN Conference on Programming Language
  Design and Implementation}, ser. PLDI ’14.\hskip 1em plus 0.5em minus
  0.4em\relax New York, NY, USA: Association for Computing Machinery, 2014, p.
  259–269.

\bibitem{10.5555/1924943.1924971}
W.~Enck, P.~Gilbert, B.-G. Chun, L.~P. Cox, J.~Jung, P.~McDaniel, and A.~N.
  Sheth, ``Taintdroid: An information-flow tracking system for realtime privacy
  monitoring on smartphones,'' in \emph{Proceedings of the 9th USENIX
  Conference on Operating Systems Design and Implementation}, ser.
  OSDI’10.\hskip 1em plus 0.5em minus 0.4em\relax USA: USENIX Association,
  2010, p. 393–407.

\bibitem{8010886}
W.~{You}, B.~{Liang}, W.~{Shi}, P.~{Wang}, and X.~{Zhang}, ``Taintman: An
  art-compatible dynamic taint analysis framework on unmodified and non-rooted
  android devices,'' \emph{IEEE Transactions on Dependable and Secure
  Computing}, vol.~17, no.~1, pp. 209--222, Jan 2020.

\bibitem{doi:10.1002/spe.2346}
R.~Pawlak, M.~Monperrus, N.~Petitprez, C.~Noguera, and L.~Seinturier, ``Spoon:
  A library for implementing analyses and transformations of java source
  code,'' \emph{Software: Practice and Experience}, vol.~46, no.~9, pp.
  1155--1179, 2016.

\bibitem{appium}
\BIBentryALTinterwordspacing
Appium. [Online]. Available: \url{http://appium.io/}
\BIBentrySTDinterwordspacing

\bibitem{jacoco}
\BIBentryALTinterwordspacing
Jacoco java code coverage library. [Online]. Available:
  \url{https://www.eclemma.org/jacoco/}
\BIBentrySTDinterwordspacing

\bibitem{10.1145/3106237.3106244}
R.~Jabbarvand and S.~Malek, ``Mudroid: An energy-aware mutation testing
  framework for android,'' in \emph{Proceedings of the 2017 11th Joint Meeting
  on Foundations of Software Engineering}, ser. ESEC/FSE 2017.\hskip 1em plus
  0.5em minus 0.4em\relax New York, NY, USA: Association for Computing
  Machinery, 2017, p. 208–219.

\bibitem{10.1145/2789168.2790107}
X.~Chen, A.~Jindal, N.~Ding, Y.~C. Hu, M.~Gupta, and R.~Vannithamby,
  ``Smartphone background activities in the wild: Origin, energy drain, and
  optimization,'' in \emph{Proceedings of the 21st Annual International
  Conference on Mobile Computing and Networking}, ser. MobiCom '15.\hskip 1em
  plus 0.5em minus 0.4em\relax New York, NY, USA: Association for Computing
  Machinery, 2015, p. 40–52.

\bibitem{10.1145/3064176.3064206}
N.~Ding and Y.~C. Hu, ``Gfxdoctor: A holistic graphics energy profiler for
  mobile devices,'' in \emph{Proceedings of the Twelfth European Conference on
  Computer Systems}, ser. EuroSys ’17.\hskip 1em plus 0.5em minus 0.4em\relax
  New York, NY, USA: Association for Computing Machinery, 2017, p. 359–373.

\bibitem{10.1145/2736277.2741635}
A.~Nika, Y.~Zhu, N.~Ding, A.~Jindal, Y.~C. Hu, X.~Zhou, B.~Y. Zhao, and
  H.~Zheng, ``Energy and performance of smartphone radio bundling in outdoor
  environments,'' in \emph{Proceedings of the 24th International Conference on
  World Wide Web}, ser. WWW '15.\hskip 1em plus 0.5em minus 0.4em\relax
  Republic and Canton of Geneva, CHE: International World Wide Web Conferences
  Steering Committee, 2015, p. 809–819.

\bibitem{10.1145/2307636.2307658}
J.~Huang, F.~Qian, A.~Gerber, Z.~M. Mao, S.~Sen, and O.~Spatscheck, ``A close
  examination of performance and power characteristics of 4g lte networks,'' in
  \emph{Proceedings of the 10th International Conference on Mobile Systems,
  Applications, and Services}, ser. MobiSys '12.\hskip 1em plus 0.5em minus
  0.4em\relax New York, NY, USA: Association for Computing Machinery, 2012, p.
  225–238.

\bibitem{180294}
F.~Xu, Y.~Liu, Q.~Li, and Y.~Zhang, ``V-edge: Fast self-constructive power
  modeling of smartphones based on battery voltage dynamics,'' in \emph{10th
  {USENIX} Symposium on Networked Systems Design and Implementation ({NSDI}
  13)}.\hskip 1em plus 0.5em minus 0.4em\relax Lombard, IL: {USENIX}
  Association, Apr. 2013, pp. 43--55.

\bibitem{6888881}
L.~{Sun}, R.~K. {Sheshadri}, W.~{Zheng}, and D.~{Koutsonikolas}, ``Modeling
  wifi active power/energy consumption in smartphones,'' in \emph{2014 IEEE
  34th International Conference on Distributed Computing Systems}, 2014, pp.
  41--51.

\bibitem{10.1145/1999995.2000026}
F.~Qian, Z.~Wang, A.~Gerber, Z.~Mao, S.~Sen, and O.~Spatscheck, ``Profiling
  resource usage for mobile applications: A cross-layer approach,'' in
  \emph{Proceedings of the 9th International Conference on Mobile Systems,
  Applications, and Services}, ser. MobiSys '11.\hskip 1em plus 0.5em minus
  0.4em\relax New York, NY, USA: Association for Computing Machinery, 2011, p.
  321–334.

\bibitem{2020ftrace}
\BIBentryALTinterwordspacing
(2020) Using ftrace. [Online]. Available:
  \url{https://source.android.com/devices/tech/debug/ftrace}
\BIBentrySTDinterwordspacing

\bibitem{10.1145/2568225.2568321}
D.~Li, A.~H. Tran, and W.~G.~J. Halfond, ``Making web applications more energy
  efficient for oled smartphones,'' in \emph{Proceedings of the 36th
  International Conference on Software Engineering}, ser. ICSE 2014.\hskip 1em
  plus 0.5em minus 0.4em\relax New York, NY, USA: Association for Computing
  Machinery, 2014, p. 527–538.

\bibitem{martin2000design}
R.~C. Martin, ``Design principles and design patterns,'' \emph{Object Mentor},
  vol.~1, no.~34, p. 597, 2000.

\bibitem{7884614}
A.~Carette, M.~A.~A. Younes, G.~Hecht, N.~Moha, and R.~Rouvoy, ``Investigating
  the energy impact of android smells,'' in \emph{2017 IEEE 24th International
  Conference on Software Analysis, Evolution and Reengineering (SANER)}, 2017,
  pp. 115--126.

\bibitem{cruz2019catalog}
L.~Cruz and R.~Abreu, ``Catalog of energy patterns for mobile applications,''
  \emph{Empirical Software Engineering}, vol.~24, no.~4, pp. 2209--2235, 2019.

\bibitem{9054858}
M.~Couto, J.~Saraiva, and J.~P. Fernandes, ``Energy refactorings for android in
  the large and in the wild,'' in \emph{2020 IEEE 27th International Conference
  on Software Analysis, Evolution and Reengineering (SANER)}, 2020, pp.
  217--228.

\bibitem{10.1145/2528265.2528267}
S.~Apel, S.~Kolesnikov, N.~Siegmund, C.~K\"{a}stner, and B.~Garvin, ``Exploring
  feature interactions in the wild: The new feature-interaction challenge,'' in
  \emph{Proceedings of the 5th International Workshop on Feature-Oriented
  Software Development}, ser. FOSD '13.\hskip 1em plus 0.5em minus 0.4em\relax
  New York, NY, USA: Association for Computing Machinery, 2013, p. 1–8.

\bibitem{kolesnikov_relation_2019}
S.~Kolesnikov, N.~Siegmund, C.~Kästner, and S.~Apel, ``On the relation of
  control-flow and performance feature interactions: a case study,''
  \emph{Empirical Software Engineering}, vol.~24, no.~4, pp. 2410--2437, Aug.
  2019.

\bibitem{10.1145/3173162.3173206}
S.~Wang, C.~Li, H.~Hoffmann, S.~Lu, W.~Sentosa, and A.~I. Kistijantoro,
  ``Understanding and auto-adjusting performance-sensitive configurations,'' in
  \emph{Proceedings of the Twenty-Third International Conference on
  Architectural Support for Programming Languages and Operating Systems}, ser.
  ASPLOS ’18.\hskip 1em plus 0.5em minus 0.4em\relax New York, NY, USA:
  Association for Computing Machinery, 2018, p. 154–168.

\bibitem{kitchenham2008personal}
B.~A. Kitchenham and S.~L. Pfleeger, ``Personal opinion surveys,'' in
  \emph{Guide to advanced empirical software engineering}.\hskip 1em plus 0.5em
  minus 0.4em\relax Springer, 2008, pp. 63--92.

\bibitem{nagappan2013diversity}
M.~Nagappan, T.~Zimmermann, and C.~Bird, ``Diversity in software engineering
  research,'' in \emph{Proceedings of the 2013 9th joint meeting on foundations
  of software engineering}, 2013, pp. 466--476.

\bibitem{10.1145/3379469}
H.-C. Kuo, J.~Chen, S.~Mohan, and T.~Xu, ``Set the configuration for the heart
  of the os: On the practicality of operating system kernel debloating,''
  \emph{Proc. ACM Meas. Anal. Comput. Syst.}, vol.~4, no.~1, May 2020.

\bibitem{8952456}
L.~{Bao}, X.~{Liu}, F.~{Wang}, and B.~{Fang}, ``Actgan: Automatic configuration
  tuning for software systems with generative adversarial networks,'' in
  \emph{2019 34th IEEE/ACM International Conference on Automated Software
  Engineering (ASE)}, 2019, pp. 465--476.

\bibitem{10.1145/2739480.2754648}
F.~Wu, W.~Weimer, M.~Harman, Y.~Jia, and J.~Krinke, ``Deep parameter
  optimisation,'' in \emph{Proceedings of the 2015 Annual Conference on Genetic
  and Evolutionary Computation}, ser. GECCO '15.\hskip 1em plus 0.5em minus
  0.4em\relax New York, NY, USA: Association for Computing Machinery, 2015, p.
  1375–1382.

\bibitem{10.1145/2714064.2660196}
M.~Kambadur and M.~A. Kim, ``An experimental survey of energy management across
  the stack,'' \emph{SIGPLAN Not.}, vol.~49, no.~10, p. 329–344, Oct. 2014.

\bibitem{246156}
Z.~Cao, G.~Kuenning, and E.~Zadok, ``Carver: Finding important parameters for
  storage system tuning,'' in \emph{18th {USENIX} Conference on File and
  Storage Technologies ({FAST} 20)}.\hskip 1em plus 0.5em minus 0.4em\relax
  Santa Clara, CA: {USENIX} Association, Feb. 2020, pp. 43--57.

\bibitem{215953}
Z.~Cao, V.~Tarasov, S.~Tiwari, and E.~Zadok, ``Towards better understanding of
  black-box auto-tuning: A comparative analysis for storage systems,'' in
  \emph{2018 {USENIX} Annual Technical Conference ({USENIX} {ATC} 18)}.\hskip
  1em plus 0.5em minus 0.4em\relax Boston, MA: {USENIX} Association, Jul. 2018,
  pp. 893--907.

\bibitem{10.1145/3342195.3387526}
H.-C. Kuo, D.~Williams, R.~Koller, and S.~Mohan, ``A linux in unikernel
  clothing,'' in \emph{Proceedings of the Fifteenth European Conference on
  Computer Systems}, ser. EuroSys ’20.\hskip 1em plus 0.5em minus 0.4em\relax
  New York, NY, USA: Association for Computing Machinery, 2020.

\bibitem{234950}
A.~Mahgoub, P.~Wood, A.~Medoff, S.~Mitra, F.~Meyer, S.~Chaterji, and S.~Bagchi,
  ``{SOPHIA}: Online reconfiguration of clustered nosql databases for
  time-varying workloads,'' in \emph{2019 {USENIX} Annual Technical Conference
  ({USENIX} {ATC} 19)}.\hskip 1em plus 0.5em minus 0.4em\relax Renton, WA:
  {USENIX} Association, Jul. 2019, pp. 223--240.

\bibitem{10.1145/3126908.3126951}
Y.~Li, K.~Chang, O.~Bel, E.~L. Miller, and D.~D.~E. Long, ``Capes: Unsupervised
  storage performance tuning using neural network-based deep reinforcement
  learning,'' in \emph{Proceedings of the International Conference for High
  Performance Computing, Networking, Storage and Analysis}, ser. SC
  ’17.\hskip 1em plus 0.5em minus 0.4em\relax New York, NY, USA: Association
  for Computing Machinery, 2017.

\bibitem{10.1145/2786805.2786845}
N.~Siegmund, A.~Grebhahn, S.~Apel, and C.~K\"{a}stner, ``Performance-influence
  models for highly configurable systems,'' in \emph{Proceedings of the 2015
  10th Joint Meeting on Foundations of Software Engineering}, ser. ESEC/FSE
  2015.\hskip 1em plus 0.5em minus 0.4em\relax New York, NY, USA: Association
  for Computing Machinery, 2015, p. 284–294.

\bibitem{10.1145/3236024.3236074}
P.~Jamshidi, M.~Velez, C.~K\"{a}stner, and N.~Siegmund, ``Learning to sample:
  Exploiting similarities across environments to learn performance models for
  configurable systems,'' in \emph{Proceedings of the 2018 26th ACM Joint
  Meeting on European Software Engineering Conference and Symposium on the
  Foundations of Software Engineering}, ser. ESEC/FSE 2018.\hskip 1em plus
  0.5em minus 0.4em\relax New York, NY, USA: Association for Computing
  Machinery, 2018, p. 71–82.

\bibitem{8812049}
C.~{Kaltenecker}, A.~{Grebhahn}, N.~{Siegmund}, J.~{Guo}, and S.~{Apel},
  ``Distance-based sampling of software configuration spaces,'' in \emph{2019
  IEEE/ACM 41st International Conference on Software Engineering (ICSE)}, May
  2019, pp. 1084--1094.

\bibitem{10.1145/3342195.3387554}
D.~Rogora, A.~Carzaniga, A.~Diwan, M.~Hauswirth, and R.~Soul\'{e}, ``Analyzing
  system performance with probabilistic performance annotations,'' in
  \emph{Proceedings of the Fifteenth European Conference on Computer Systems},
  ser. EuroSys ’20.\hskip 1em plus 0.5em minus 0.4em\relax New York, NY, USA:
  Association for Computing Machinery, 2020.

\bibitem{velez_configcrusher_2020}
M.~Velez, P.~Jamshidi, F.~Sattler, N.~Siegmund, S.~Apel, and C.~Kästner,
  ``{ConfigCrusher}: towards white-box performance analysis for configurable
  systems,'' \emph{Automated Software Engineering}, Aug. 2020.

\bibitem{6224257}
C.~Sahin, F.~Cayci, I.~L.~M. Gutiérrez, J.~Clause, F.~Kiamilev, L.~Pollock,
  and K.~Winbladh, ``Initial explorations on design pattern energy usage,'' in
  \emph{2012 First International Workshop on Green and Sustainable Software
  (GREENS)}, 2012, pp. 55--61.

\bibitem{10.1145/2714064.2660235}
G.~Pinto, F.~Castor, and Y.~D. Liu, ``Understanding energy behaviors of thread
  management constructs,'' \emph{SIGPLAN Not.}, vol.~49, no.~10, p. 345–360,
  Oct. 2014.

\bibitem{222559}
``Differential energy profiling: Energy optimization via diffing similar
  apps,'' in \emph{13th {USENIX} Symposium on Operating Systems Design and
  Implementation ({OSDI} 18)}.\hskip 1em plus 0.5em minus 0.4em\relax Carlsbad,
  CA: {USENIX} Association, Oct. 2018.

\bibitem{8812097}
R.~{Jabbarvand}, J.~{Lin}, and S.~{Malek}, ``Search-based energy testing of
  android,'' in \emph{2019 IEEE/ACM 41st International Conference on Software
  Engineering (ICSE)}, 2019, pp. 1119--1130.

\bibitem{10.1145/3368089.3409677}
R.~Jabbarvand, F.~Mehralian, and S.~Malek, ``Automated construction of energy
  test oracles for android,'' in \emph{Proceedings of the 28th ACM Joint
  Meeting on European Software Engineering Conference and Symposium on the
  Foundations of Software Engineering}, ser. ESEC/FSE 2020.\hskip 1em plus
  0.5em minus 0.4em\relax New York, NY, USA: Association for Computing
  Machinery, 2020, p. 927–938.

\bibitem{10.1145/3395363.3397350}
X.~Li, Y.~Yang, Y.~Liu, J.~P. Gallagher, and K.~Wu, ``Detecting and diagnosing
  energy issues for mobile applications,'' in \emph{Proceedings of the 29th ACM
  SIGSOFT International Symposium on Software Testing and Analysis}, ser. ISSTA
  2020.\hskip 1em plus 0.5em minus 0.4em\relax New York, NY, USA: Association
  for Computing Machinery, 2020, p. 115–127.

\end{thebibliography}

\end{document}